\documentclass[aps,showpacs,amsmath,amssymb,prb,groupedaddress,preprint]{revtex4-1}
\usepackage{color}
\usepackage{graphicx,dcolumn,diagbox,bm}
\usepackage[latin9]{inputenc}
\usepackage{amsmath}
\usepackage{graphics}
\usepackage{gensymb}
\usepackage{subfigure}
\usepackage{pdftexcmds}
\usepackage{ifpdf}
\usepackage{amssymb}
\usepackage{dsfont}

\newcommand{\MTX} {{\mbox{MT${}_2$X${}_4$}}}
\newcommand{\MGS} {{\mbox{MGa${}_2$S${}_4$}}}
\newcommand{\NGS} {{\mbox{NiGa${}_2$S${}_4$}}}
\newcommand{\FGS} {{\mbox{FeGa${}_2$S${}_4$}}}

\newcommand{\DSC} {({\ensuremath{\frac{d\sigma}} {d\Omega}})}
\newcommand{\half}{{\ensuremath{\frac{1}{2}}}}
\newcommand{\third}{{\ensuremath{\frac{1}{3}}}}
\newcommand{\six}{{\ensuremath{\frac{1}{6}}}}
\newcommand{\fourth}{{\ensuremath{\frac{1}{4}}}}
\definecolor{green}{rgb}{0.15,0.85,0.35}

\allowdisplaybreaks
\makeatother
 
\begin{document}

\title{Magnetic correlations in triangular antiferromagnet FeGa$_2$S$_4$}
\author{K. Guratinder}
\affiliation{Laboratory for Neutron Scattering and Imaging, Paul Scherrer Institute, CH-5232 Villigen, Switzerland}
\affiliation{Department of Quantum Matter Physics, University of Geneva, CH-1211 Geneva, Switzerland }
\author{M. Schmidt}
\affiliation{Max-Planck-Institut f\"{u}r Chemische Physik fester Stoffe, 01187 Dresden, Germany}
\author{H. C. Walker, R. Bewley}
\affiliation{Rutherford Appleton Laboratory, ISIS Facility, Chilton, Didcot, Oxon OX11 0QX, United Kingdom}
\author{M. W\"{o}rle}
\affiliation{Laboratorium f\"{u}r Anorganische Chemie, CH-8093 Z\"{u}rich, Switzerland}
\author{D. Cabra}
\affiliation{Instituto de F\'isica de L\'iquidos y Sistemas Biol\'ogicos, CCT La Plata, CONICET and Departamento de F\'isica,
Facultad de Ciencias Exactas, Universidad Nacional de La Plata, C.C. 67, 1900 La Plata, Argentina}
\author{S. A. Osorio}
\affiliation{Instituto de F\'isica de L\'iquidos y Sistemas Biol\'ogicos, CCT La Plata, CONICET and Departamento de F\'isica,
Facultad de Ciencias Exactas, Universidad Nacional de La Plata, C.C. 67, 1900 La Plata, Argentina}
\author{M. Villalba}
\affiliation{Instituto de F\'isica de L\'iquidos y Sistemas Biol\'ogicos, CCT La Plata, CONICET and Departamento de F\'isica,
Facultad de Ciencias Exactas, Universidad Nacional de La Plata, C.C. 67, 1900 La Plata, Argentina}
\author{A. K. Madsen}
\affiliation{Quantum Criticality and Dynamics Group, Paul Scherrer Institute, CH-5232 Villigen, Switzerland}
\author{L. Keller}
\affiliation{Laboratory for Neutron Scattering and Imaging, Paul Scherrer Institute, CH-5232 Villigen , Switzerland}
\author{A. Wildes}
\affiliation{Institut Laue-Langevin, 156X, 38042 Grenoble C\'{e}dex, France}
\author{P. Puphal}
\affiliation{Laboratory for Multiscale Materials Experiments, Paul Scherrer Institute, CH-5232 Villigen, Switzerland}
\author{A. Cervellino}
\affiliation{Swiss Light Source, Paul Scherrer Institute, CH-5232 Villigen, Switzerland}
\author{Ch. R\"{u}egg}
\affiliation{Laboratory for Neutron Scattering and Imaging, Paul Scherrer Institute, CH-5232 Villigen, Switzerland}
\affiliation{Department of Quantum Matter Physics, University of Geneva, CH-1211 Geneva, Switzerland}
\altaffiliation[Presently at ]{Research Division Neutrons and Muons, Paul Scherrer Institute, CH-5232 Villigen, Switzerland}
\author{O. Zaharko}\email{oksana.zaharko@psi.ch}
\affiliation{Laboratory for Neutron Scattering and Imaging, Paul Scherrer Institute, CH-5232 Villigen, Switzerland.}

\date{\today}

\begin{abstract}
The crystal structure and magnetic correlations in triangular antiferromagnet {\FGS} are studied by x-ray diffraction, magnetic susceptibility, neutron diffraction and neutron inelastic scattering. We report significant mixing at the cation sites and disentangle
magnetic properties dominated by major and minor magnetic sites. The magnetic short-range correlations at 0.77 \AA$^{-1}$ correspond to the major sites and being static at base temperature they evolve into dynamic correlations around 30 - 50 K. The minor sites contribute to the magnetic peak at 0.6 \AA$^{-1}$, which vanishes at 5.5 K.\\
Our analytical studies of triangular lattice models with bilinear and biquadratic terms provide the ratios between exchanges for the proposed ordering vectors. The modelling of the inelastic neutron spectrum within linear spin wave theory results in the set of exchange couplings $J_1=1.7$\,meV, $J_2=0.9$\,meV, $J_3=0.8$\,meV for the bilinear Heisenberg Hamiltonian. However, not all features of the excitation spectrum are explained with this model.  
\end{abstract}

\maketitle

\section{Introduction}{\label{Sec1}}

\begin{figure}
\centering 
\includegraphics[width=0.80\columnwidth]{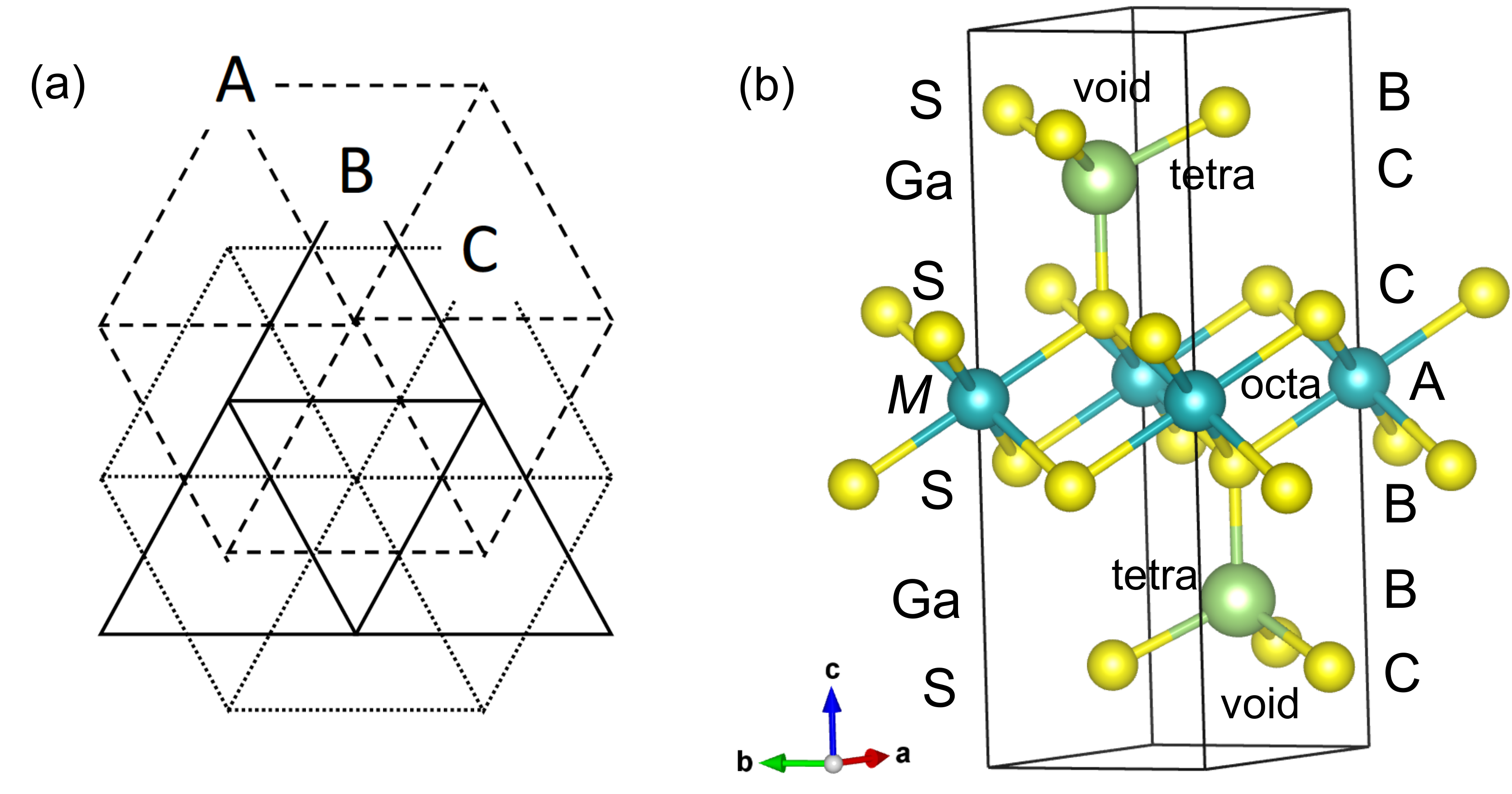} 
\caption{a) Projection of the A-, B-, C-layers on the $ab$-plane.
b) View on the {\MGS} unit cell with labeled (from left to right) sequence of S, Ga and M atoms, coordination of the voids between the S-layers and order of the A, B, C layers.}
\label{fig_layers} 
\end{figure}

The interest in the {\MTX} family of compounds (M = Ni, Fe; T = Ga, Al; X = S, Se) arises due to the quasi- 2-dimensional (2D) triangular geometry of the magnetic
M-sublattice. This family, especially {\NGS}, are strong candidates as an experimental realisation of the triangular lattice antiferromagnet (TLAFM), the extensively studied theoretical model in frustrated magnetism. 
For the spin S=1/2 TLAFM case a resonating valence bond spin liquid ground state was proposed by Anderson\cite{Anderson1973},
which was later shown instead to have a classical-like 120 $\deg$ planar ground state\cite{Jolicoeur1990}  in spite of the quantum fluctuations that only reduce the value of the order parameter. For Ising spins, on the other hand, it has been shown that the first nearest-neighbor ($J_1$) Ising TLAFM model stays disordered even down to zero temperature.
Additional exchange interactions spanning over the second ($J_2$) and third ($J_3$) nearest-neighbors could enhance frustration and for certain combinations of the exchange parameters exotic ground states are predicted. For example for $J_1/J_3$ = - {\third} or $J_1/J_2$ = - {\fourth} a skyrmion crystal multi-$k$ state should be realised\cite{Okubo2012}. Such swirling configurations with topological properties are of interest for the fast emerging field of spintronics.\\
{\MGS} compounds with M = Ni, Fe crystallise in the trigonal space group $P\bar{3}m1$\cite{Dogguy1980}. All atoms are located 
on triangular layers, which generally could be stacked in the A-, B- or C-fashion (Fig.~\ref{fig_layers}a).
In {\MGS} the sulphur layers stack in the BCBC sequence providing
octahedral and tetrahedral voids. The octahedral voids are occupied by the smaller M-cations, and the tetrahedral voids by Ga (Fig.~\ref{fig_layers}b).
The unit cell of {\MGS} is terminated by an empty void layer.
Our x-ray diffraction study detected significant mixed occupancies (inversion) of the cation sites which we associate with highly nontrivial magnetic behaviour of the {\MGS} series. This feature has not been observed before and it is one of the important outcomes of our work.\\
%
%
The published bulk properties\cite{Nakatsuji2007, Myoung2008, Zhao2012} suggest stronger exchange couplings for {\FGS} (Fe$^{2+}$, S=2) compared to {\NGS} (Ni$^{2+}$, S=1), but the main observations remain very similar.
The Curie-Weiss temperature is $\theta_{CW}$=160 K for {\FGS} ($\theta_{CW}$=80 K for the Ni-analogue) and there is a bifurcation between field-cooled (FC) and zero-field-cooled (ZFC) dc magnetic susceptibilities at T$_f$(Fe)=16 K (T$_f$(Ni)=8-9 K). No long-range order has been detected so far for both systems, but an inelastic neutron scattering (INS) study\cite{Stock2010} for {\NGS} reported incommensurate (ICM) quasielastic short-range magnetic correlations in the vicinity of Q=(0.155 0.155 0) and spin waves propagating from this Q-point up to 3.5 meV. The dynamic muon relaxation rates \cite{Reotier2012, Zhao2012} show an anomaly at T$^*$(Fe)=33 K (T$^*$(Ni)=10 K) approaching correlation times ${\tau_c}$(Fe)=10$^{-6}$ s (${\tau_c}$(Ni)=10$^{-7}$ s). These processes are much slower than the time scale of exchange interaction (10$^{-12}$ s) and the corresponding states are interpreted as a slowly fluctuating 'spin gel'. Surprisingly, no anomaly at T$_f$ was observed by $^{57}$Fe M\"{o}ssbauer, $\mu$SR and specific heat CM, while the T$^*$ anomaly is not present in magnetic bulk properties\cite{Nakatsuji2007, Myoung2008}. The magnetic component of the specific heat CM(T) has a double-peak structure for both compounds with a first maximum at 10 K for both and a second maximum at 60 K (100 K) for {\FGS}\cite{Nakatsuji2007} ({\NGS}\cite{Nakatsuji2005}). CM(T) has a T$^2$ dependence at low temperatures, which indicates gapless and linearly dispersive modes in 2D, and it is insensitive to magnetic fields up to 7 T. Our samples show the same magnetic behaviour and we complement experimental information on the {\FGS} system by neutron experiments, both diffraction and inelastic scattering.\\
From the theoretical side a number of extensions to the TLAFM model were proposed to explain the observed magnetic behaviour of the Ni- and Fe- analogues. A strong, dominant third-nearest-neighbor antiferromagnetic exchange interaction is supported by DFT calculations \cite{Mazin2007} and could explain the ICM magnetic propagation vector of {\NGS}\cite{Stock2010}. Biquadratic interactions, which would yield spin-nematic quadrupolar correlations, account for the double-peak structure in specific heat\cite{Tsunetsugu2006, Lauchli2006, Bhattacharjee2006}. The origin of the spin nematic could be the coupling of magnetic correlations to phonons observed by Raman scattering\cite{Valentine2020}. 
The $Z_2$ vortex model with binding-unbinding transition\cite{Tsunetsugu2007} could also explain the high temperature anomalies in magnetic properties, while linearly dispersing Halperin-Saslow modes in the absence of long-range magnetic order could elucidate the anomalous low-temperature properties\cite{Podolsky2009}. We scrutinize the TLAFM model with $J_1$, $J_2$, $J_3$ billinear exchanges and show that it is not sufficient to model our INS data. We associate the unconventional magnetic properties with the non-ideal structure of the {\MGS} family.  

\section{Crystal structure of F\lowercase{e}G\lowercase{a}$_2$S$_4$ and N\lowercase{i}G\lowercase{a}$_2$S$_4$}{\label{Sec2}}
\subsection{Study of structure imperfection by X-ray diffraction}{\label{Sec2Sub1}}
%
First we check thoroughly the crystal structure of our polycrystalline {\FGS} sample\cite{note1} analysing synchrotron x-ray diffraction powder pattern collected at 300 K. These data were well refined within the crystal structure published by Dogguy Simiri\cite{Dogguy1980}: the Ga and two S atoms located at the 2(d) sites and Fe atoms at the 1(b) site of the $P\bar{3}m1$ space group (R$_{Bragg}$=3.64 \%, R$_f$=1.99 \%, ${\chi^2}$=47.3). The fits became even better (R$_{Bragg}$=2.98 \%, R$_f$=1.89 \%, ${\chi^2}$=46.2) when the inversion between the Fe and Ga atoms was introduced with the total occupancy constrained to the stoichiometric composition {\FGS}.
According to this refinement the 1(b) site is occupied 20\% by gallium and 80\% by iron, whereas the 2(d) site - by 10\% iron and 90\% gallium, respectively.
As the model with partial inversion is only nominally better and a small amount of Ga$_2$O$_3$ impurity phase was detected in the powder pattern, we verified this result by x-ray diffraction on single crystals.\\
\begin{figure}
\centering 
\includegraphics[width=0.99\columnwidth]{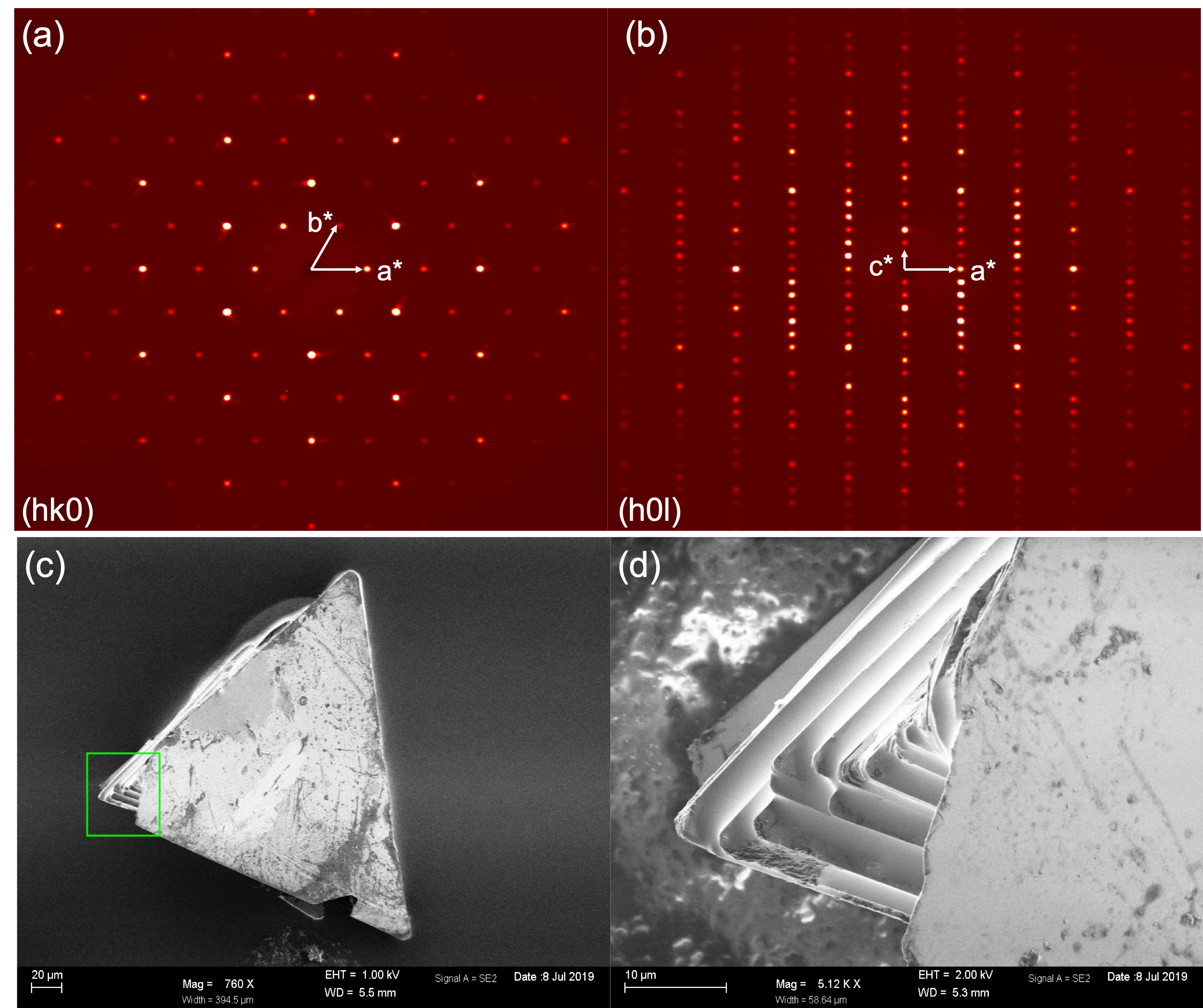} 
\caption{Top: The hk0 (left) and h0l (right) reciprocal layers reconstruction based on the X-ray single crystal diffraction of {\FGS}.
Bottom: (left) scanning electron microscopy image of a single crystal of {\FGS} and (right) zoom into its left marked corner.}
\label{fig_diffXtal}
\end{figure}
%
For several studied crystals (even very small ones) the shape of reflections was not exceptionally good - they had 'tails' typical for 
small arbitrary rotations of macroscopic domains around the [001] axis, as can be seen in the reciprocal hk0-layer reconstruction presented in
Fig.~\ref{fig_diffXtal} (top). Nevertheless, the consequent data reduction lead to good data with R$_{int}$=3.69~\% for 520 unique reflections for the Laue group $\bar{3}m1$. A refinement assuming the non-disordered {\FGS} structure with full occupancies converged at acceptable R-values:  R$_1$ = 2.83 \%, wR$_2$ = 8.21 \%, GooF = 1.184  for 514 $F_o>$ 4 $\sigma F_o$. The difference Fourier maps, as in the case of powder pattern, contained relatively strong residual electron density peaks: positive at the 1(b) Fe site (corresponding to around 3 e/\AA$^3$) and negative (corresponding to around -1.7 e/\AA$^3$) in the vicinity of the 2(d) Ga sites. Along with this no significant electron density occurred between the atomic positions of the ideal structure. The residual electron density could be accounted for by the partial inversion at the 1(b) and 2(d) sites. Refinement converged to
 21.5(3)\% Ga (and 78.4(3)\% Fe) on the 1(b) Wyckoff position and 10.8(1)\% Fe (and 89.2(1)\% Ga) on the Wyckoff position 2(d) with improved figure of merit: R$_1$ = 2.04 \%, Rw$_2$ = 6.08 \%, GooF = 1.367.\\
We took the indication of partial inversion from x-ray diffraction with great caution. Bulk properties (magnetic susceptibility, specific heat) measured by us and by other groups are very similar suggesting that partial inversion might improve the understanding of magnetic properties of the whole {\MGS} family. To clarify this, we synthesised and characterised isostructural {\NGS}. Refinement of {\NGS} single crystal x-ray diffraction data revealed the same degree of inversion as for {\FGS}. This complies with nuclear quadrupole resonance (NQR) studies of polycrystalline {\NGS} samples \cite{Takeya2008, Nambu2009}, which reported two Ga signals. Our results imply that these NQR signals could arise from gallium atoms located on two different crystallographic sites. 
We also considered whether other imperfections discussed in the literature could explain our diffraction data. For example, high-resolution electron microscopy detected the presence of the stacking faults along the [001]-direction\cite{Nambu2009}. Any significant fraction of stacking faults generated by a shift of the layers, as realised for example in the case of the rhombohedral structure of MgAl$_2$S$_4$ or by introducing layers found in Fe$_2$Ga$_2$S$_5$ or in GaS with the hexagonal  $\beta$-InSe structure, would lead to residual electron density off the atomic positions of the ideal structure. The same holds for the stacking faults along the slip planes (11-20), (10-11) and (11-23), which sometimes occur in the hexagonal close packing structures. Diffuse scattering should also be observed, but this is not the case in our diffraction patterns presented in Fig.~\ref{fig_diffXtal}, top.
We also checked for twinning, finding no indication of twofold rotations around [110], [1-10] and [001], the usual twin laws occurring in trigonal systems. Our scanning electron microscopy images clearly show that crystals consist of layers with a thickness on the micrometer scale, which are rotated against each other by small angles (Fig.~\ref{fig_diffXtal} bottom). 
One more sample imperfection was reported by Nambu et al. \cite{Nambu2009} based on energy-dispersive X-ray spectroscopy (EDXS) - nonstoichiometry in the sulfur content. Our EDXS analysis show no sulfur deficiency within the detection limit. \\
Thus we conclude that polycrystalline and single crystal {\MGS} samples have significant inversion on the cation sites. In the following text we correlate this partial inversion with magnetic bulk properties and neutron scattering results, labeling the 1(b) site as the 'major'-site and the 2(d) site as the 'minor'-site, since they are occupied by 80\% and 10\% with the magnetic ion, respectively.\\
%
\subsection{Temperature evolution of crystal structure}
\begin{figure}[ht]
\centering 
\includegraphics[width=0.99\columnwidth]{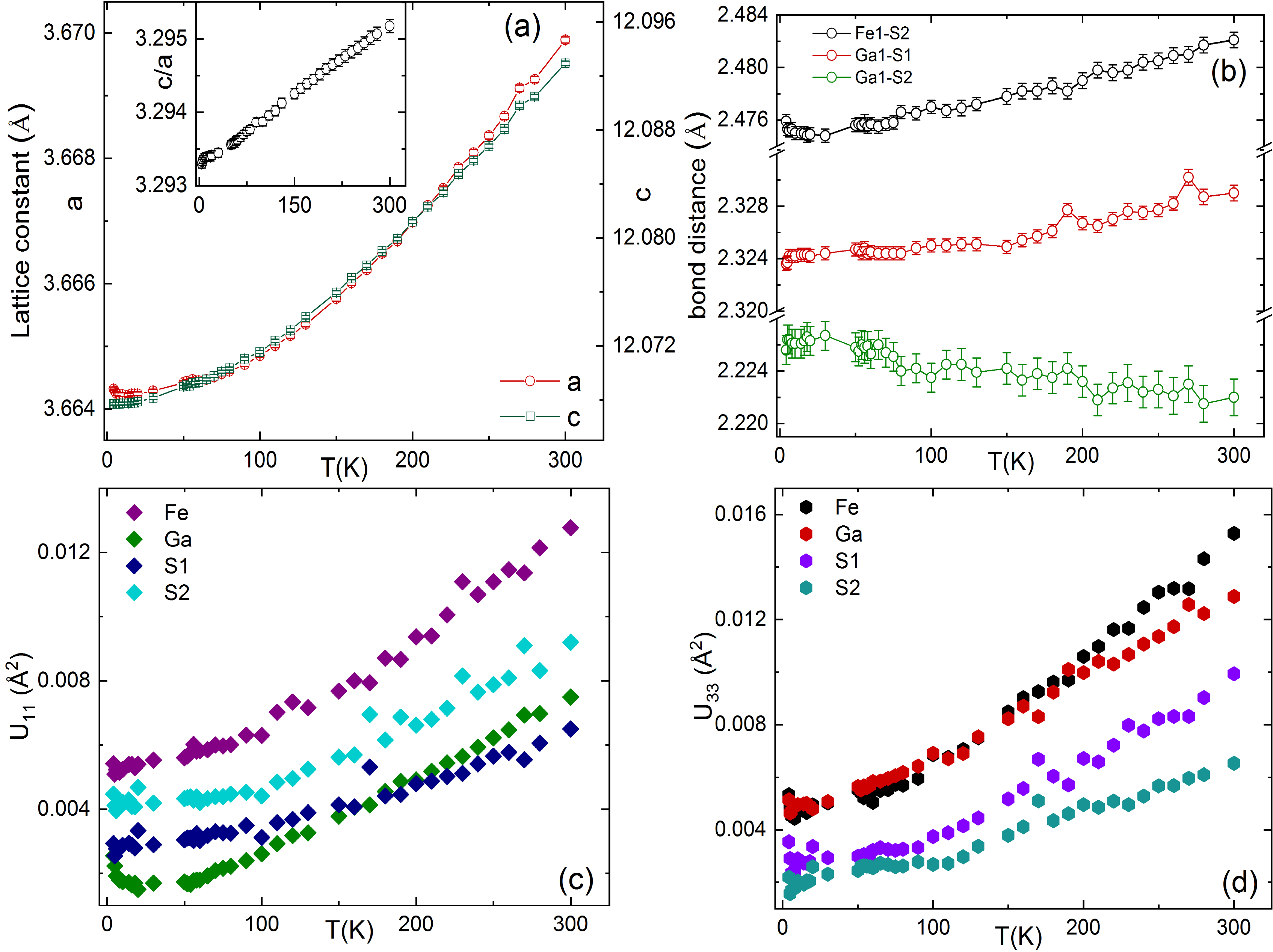}
\caption{Temperature evolution of the lattice constants (a), selected bond distances (b) and thermal parameters (c, d) of {\FGS} from synchrotron x-ray powder diffraction.}
\label{fig_lc} 
\end{figure}

{\label{Sec2Sub2}}
%
To identify possible changes to the crystal structure of {\FGS} we measured and refined synchrotron x-ray powder diffraction patterns of  a polycrystalline sample in the 5 K- 300 K temperature range. The obtained temperature evolution of the lattice constants, atomic positions and thermal parameters is presented in (Fig.~\ref{fig_lc}). 
No significant structural anomalies could be detected in these X-ray patterns. The lattice constants show a typical Debye thermal expansion, the slope of the ratio c/a changes below 50 K and near 10 K. The interatomic distances d$_{FeS2}$, d$_{GaS1}$ decrease, while d$_{GaS2}$ increases with lower temperature. The anisotropic thermal parameters of all atoms decrease very much alike, only the thermal parameters $\beta_{11}$ and $\beta_{3}$ of the 2d site decrease faster and slower, respectively.\\
These parameters of {\FGS} do not point to anomalies in vibrations of the sulfur atom S2 detected in the sister compound {\NGS} by Raman scattering\cite{Valentine2020}.
%
%
\section{Magnetic properties}{\label{Sec3}}
\subsection{Magnetic bulk properties}{\label{Sec3Sub1}}
\begin{figure}
\centering 
\includegraphics[width=0.99\columnwidth]{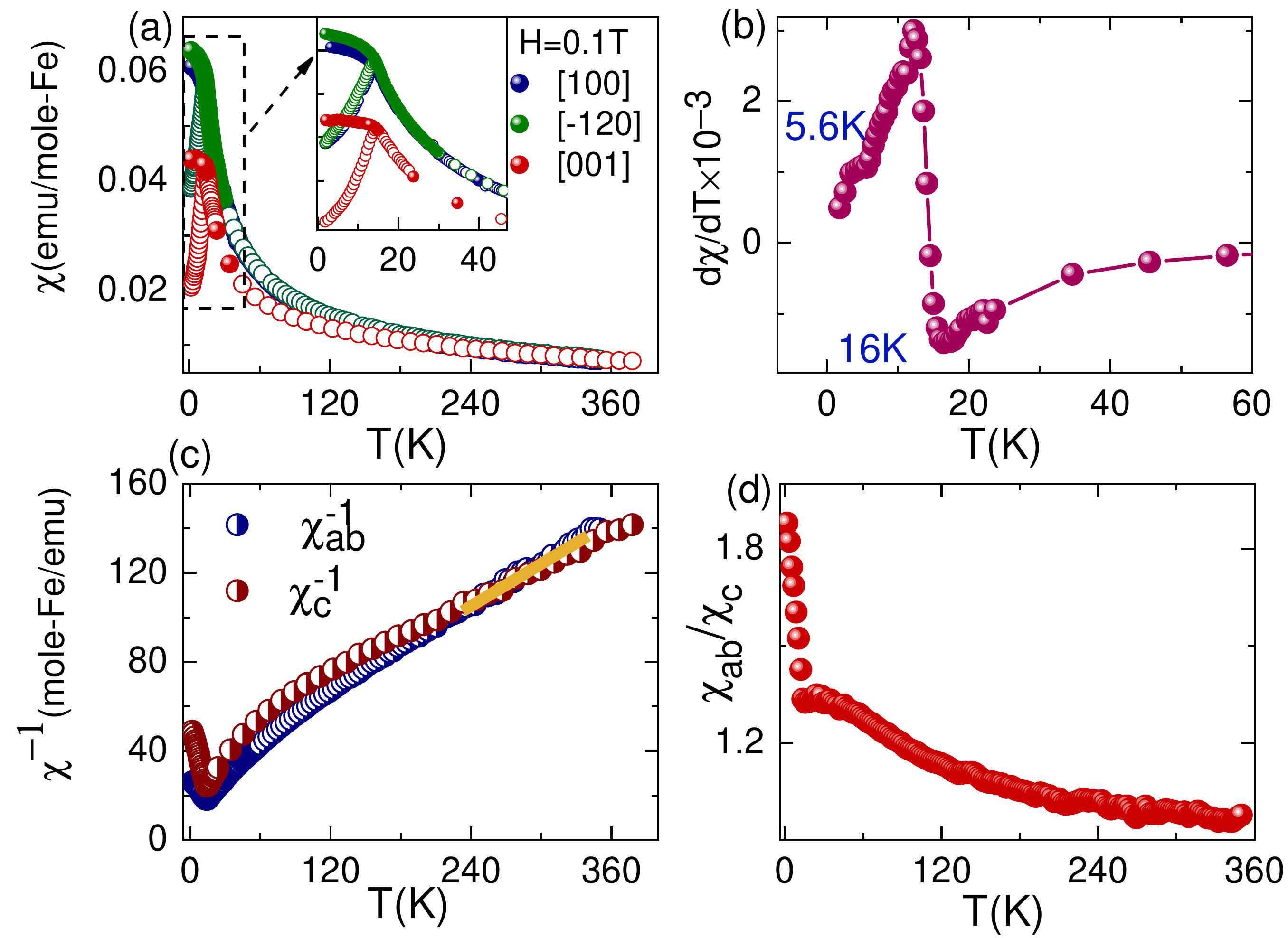} 
\caption{a) Magnetic susceptibility $\chi$ measured at 0.1 T on a single crystal of {\FGS} (mass m=2.87 mg) along the three principal directions $[100]$, $[-120]$ and $[001]$. Zero field cooled (ZFC) and field cooled (FC) $\chi$ are plotted by empty and filled circles, respectively. Inset:  zoom of the low temperature region, b) derivative of ZFC magnetic susceptibility highlighting deflection points. c) Inverse susceptibility ${1/\chi}$ for H$//[100]$ (blue) and H$//[001]$ (red) and fit  to the Curie-Weiss law in yellow. d) - the ratio of ZFC $\chi_{ab}/\chi_c$.}
\label{fig_susc} 
\end{figure}
%
We briefly report magnetic bulk properties of our {\FGS} samples. Fig.~\ref{fig_susc} (left) presents the magnetic susceptibility of a {\FGS} single crystal measured at 0.1 T in the range 1.8 K - 350 K for the three principal orthogonal crystallographic directions: $[001]$, $[100]$ and $[-120]$. At high temperatures (200 - 350 K) the curves almost overlap and fits to the Curie-Weiss law yield antiferromagnetic Curie-Weiss temperatures $\Theta_{\rm CW}$= -165(11) K  for the $[001]$ direction and -151(9) K for $[100]$ and $[-120]$ and effective magnetic moments $m_{eff}$=5.53 $\mu_B$ and 5.08 $\mu_B$, respectively. The susceptibility in the $ab$-plane and along the $c$-axis deviate below 200 K and this anisotropy reaches 20\% near 16 K. At T$_f$=16 K the susceptibility has a kink and a bifurcation between the field-cooled (FC) and zero-field- cooled (ZFC) sample for all measured directions. The difference between the FC and ZFC curves is tiny, 0.004 $m_B$/Fe. Careful inspection of the derivative of ${\chi}$ (Fig.~\ref{fig_susc}, right inset) reveals an additional change of the slope at 5.6 K. We associate it with long-range ordering of the minor Fe-spins by powder neutron diffraction (see Section \ref{Sec4}).\\
For our polycrystalline {\FGS} sample the Curie-Weiss temperature is -158.0(9) K. This results in the effective exchange constant 
\begin{equation*}
J/{k_B} =  - \frac{3 \Theta_{\rm CW}}{z S(S+1)}\sim 13~ K \sim 1.1~ meV,
\end{equation*}
here $k_B$ is the Boltzmann constant, $z = 6$ is the coordination number and $S=2$ is the spin of the Fe$^{2+}$-ions.\\
%
\subsection{Magnetic static correlations from neutron diffraction}{\label{Sec53Sub2}}
\begin{figure}
\centering 
\includegraphics[width=0.65\columnwidth]{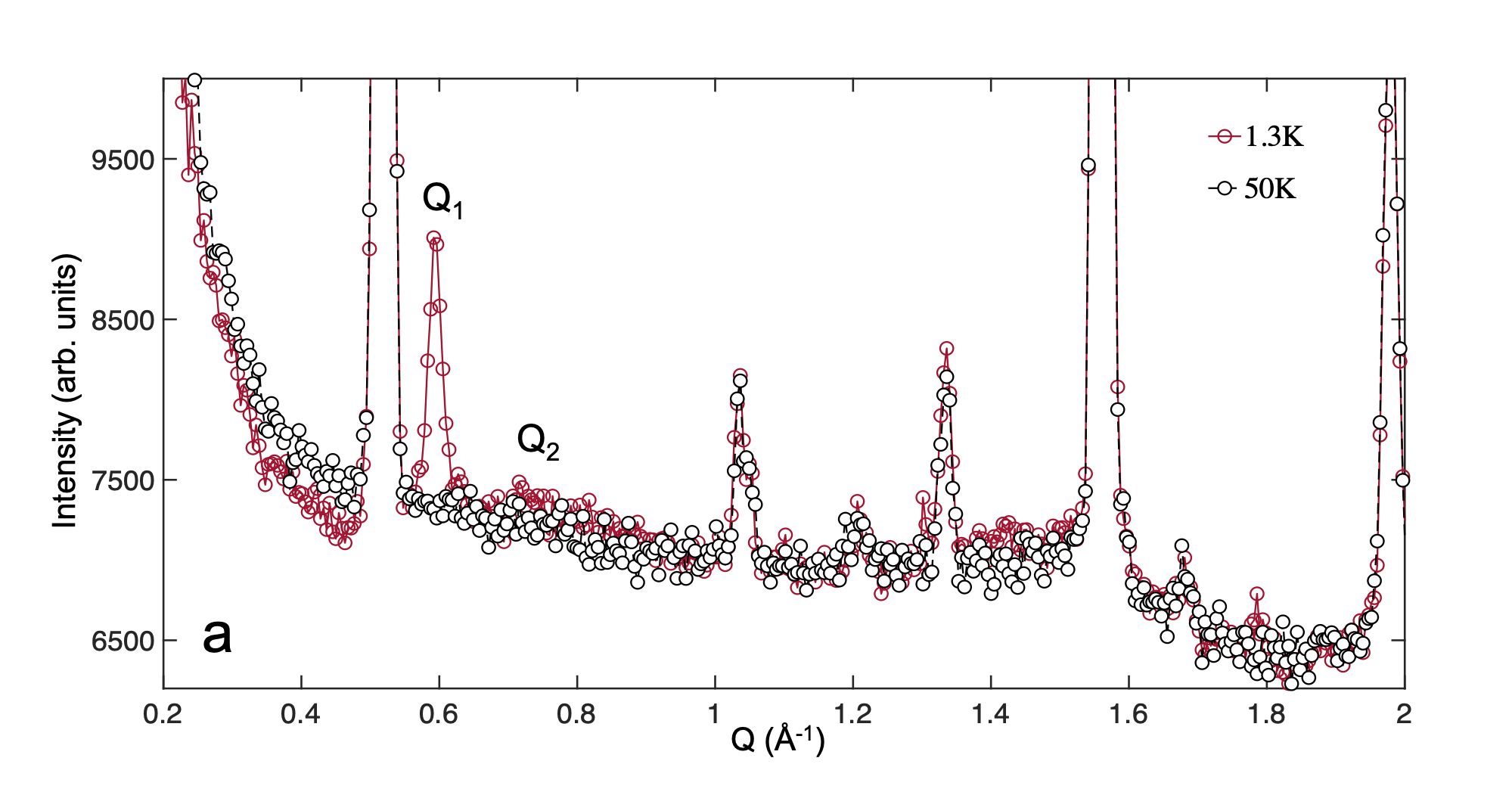} 
\includegraphics[width=0.34\columnwidth]{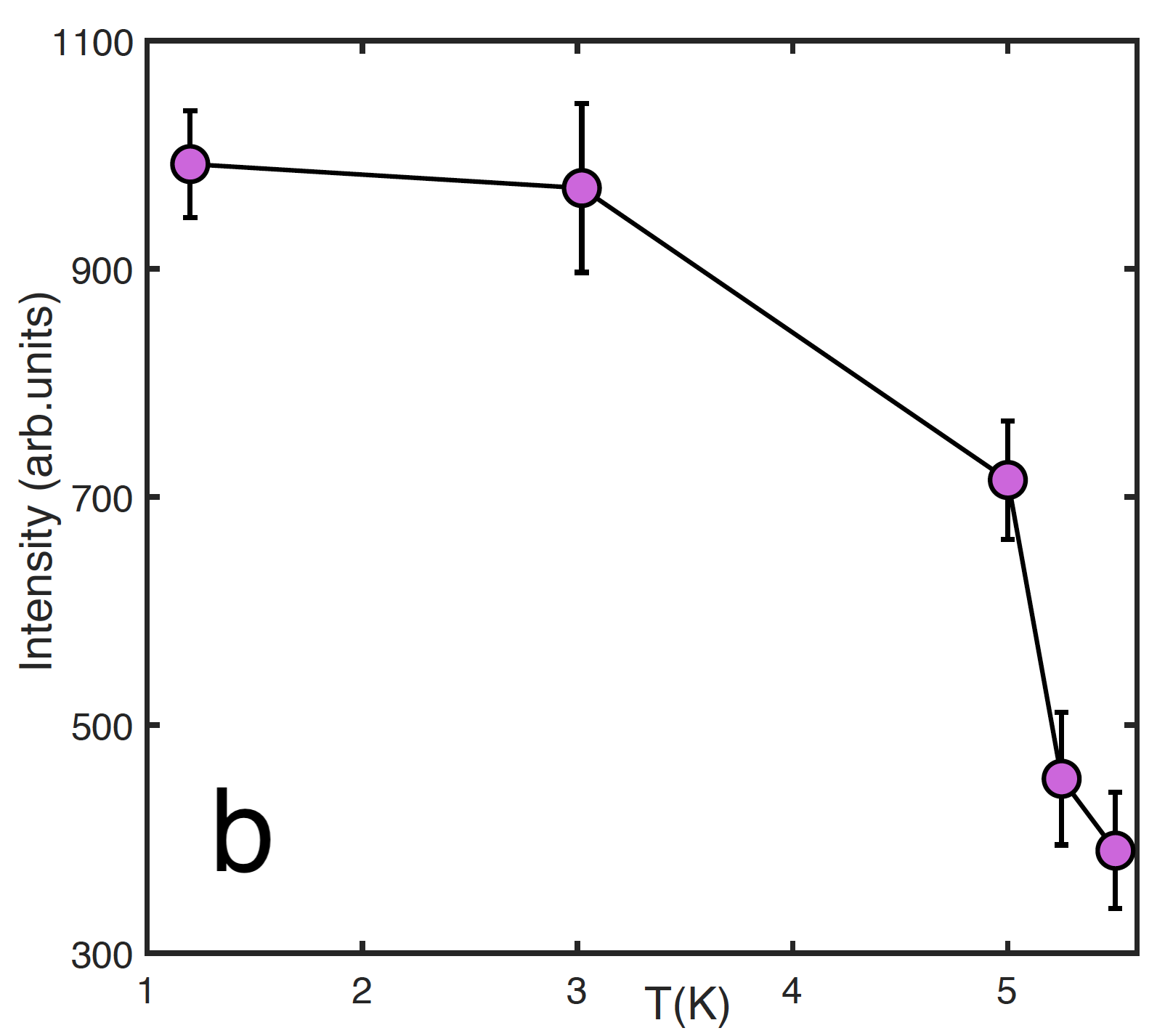} 
\caption{Neutron diffraction DMC patterns of {\FGS} at 1.3 K and 50 K (a). Temperature dependence of the magnetic peak at Q$_1$=0.6  \AA$^{-1}$ (b).}
\label{fig_nd} 
\end{figure}
\begin{figure}
\centering 
\includegraphics[width=0.49\columnwidth]{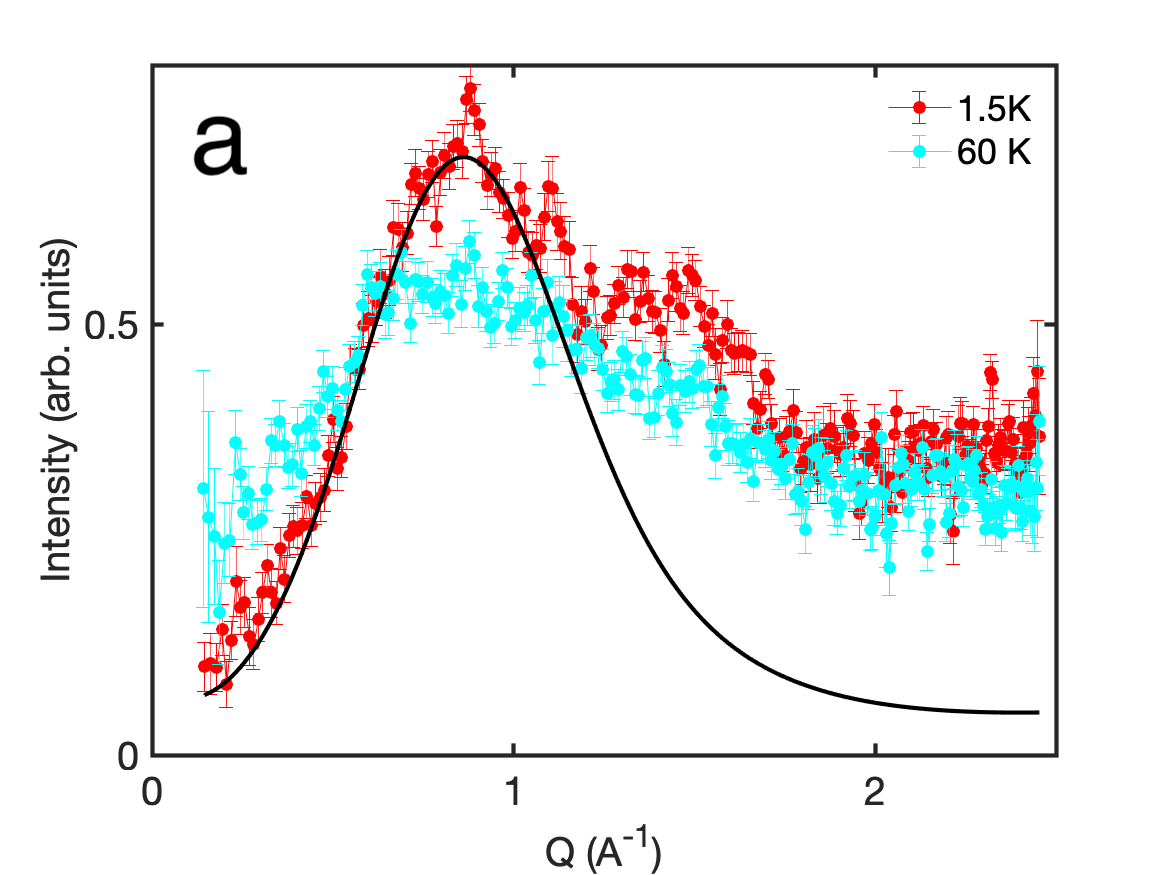} 
\includegraphics[width=0.49\columnwidth]{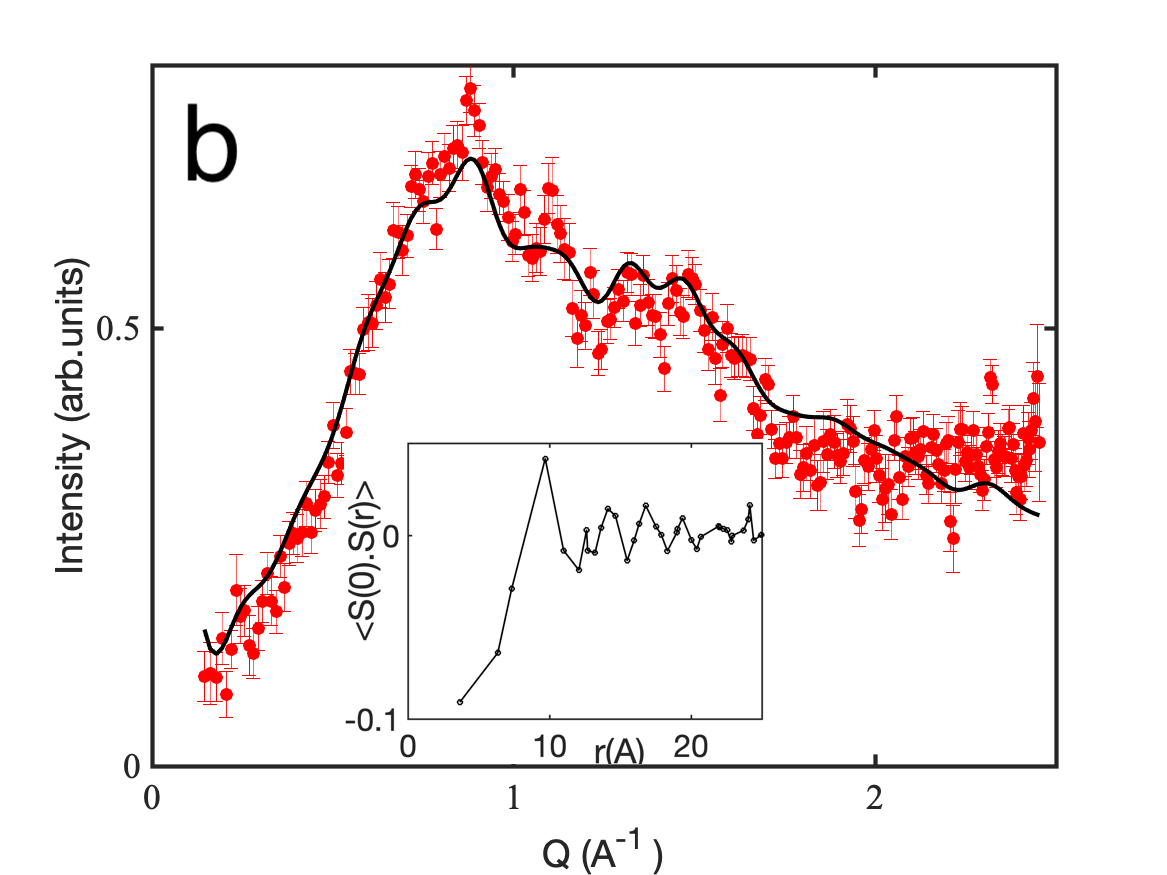} 
\caption{Magnetic diffuse scattering measured at 1.5 K (red) and 60 K (cyan) on D7 (E$_i$=3.8 meV) and fit to the Warren function a) and calculated by the RMC method (b) with resulting radial spin-correlation function $<S(0)\cdot S(r)>$ shown in insert.}
\label{fig_d7} 
\end{figure}
It is crucial to understand, if the magnetic properties of {\FGS} samples could be disentangled  into major- and minor- sites properties. First we address this question with powder neutron diffraction (PND).\\
In PND patterns collected on DMC two different features of magnetic origin appear. Prominent in Fig.~\ref{fig_nd}a is the peak at Q$_1$=0.6 \AA$^{-1}$ below T$_N$=5.5 K. It could be indexed with the incommensurate propagation vector $\bf{k}$=(0.1737(1) 0.1737(1) 0) in analogy to $\bf{k}$=(0.153 0.153 0) reported for {\NGS}. However, no long-range magnetic arrangement akin to only the major site reproduces the diffraction pattern. Models with Fe only on the 1(b) site require the second peak at (0.1737~ 0.1737~ 1) with Q=0.79 \AA$^{-1}$  to have significant intensity.
The single (0.1737~ 0.1737~ 0) peak can be reproduced by a model with the Fe moments on the 2(d) site. The moments are sinusoidally modulated along the $z$-axis. The moment amplitude  7.0(2) $\mu_B$/Fe (maximal value of the moment) would be however unreasonably large for the 10 \% occupied 2(d) site. So the M$_z$ component on  the minor 2(d) site is most probably significant but not the only 'participant' in the magnetic ordering at T$_N$=5.5 K and Q$_1$=0.6 \AA$^{-1}$.\\
The second, less evident feature in PND is diffuse magnetic scattering centered at Q$_2$=0.77 \AA$^{-1}$ (Fig.~\ref{fig_nd}a). The XYZ-polarization analysis of powder neutron diffraction patterns collected on D7
verifies that this is purely magnetic scattering (Fig.~\ref{fig_d7}). In {\NGS} a similar diffuse bump was reported beneath the $\bf{k}$=(0.153 0.153 0) magnetic peak (Fig.~4 in Ref. $[$\onlinecite{Nakatsuji2005}$]$) and it was interpreted in terms of 2D correlations extending within 25(3) \AA.
In the case of {\FGS} the diffuse feature is broader. To estimate the 2D correlation length we first used the Warren function\cite{Warren1941} initially developed to describe scattered intensity from parallel, equidistant layers randomly rotated along the normal direction.
Modelling of the magnetic PND pattern presented in Fig.~\ref{fig_d7} (a) suggests an in-plane correlation length of 10 \AA, which does not exceed the 4th nearest neighbor Fe-Fe distance.
As an alternative route, we used the reverse Monte-Carlo (RMC) algorithm implemented in the SPINVERT program\cite{Paddison2013}. We fit the magnetic PND pattern to a large array (30x30x10 unit cells) of spin vectors located on the major 1(b) sites (Fig.~\ref{fig_d7} b). The spin-pair correlation function in real space $<\bf{S}$(0)$\cdot $ $\bf{S}$(r)$>$ is calculated from the fitted spin configurations by the program SPINCORREL\cite{Paddison2013}. The negative sign of correlations for the in-plane distances d$_1^{1b}$=3.66 \AA, d$_2^{1b}$=6.35 \AA, d$_3^{1b}$=7.33 \AA~(three first points in Fig.~\ref{fig_d7}b insert) suggests antiparallel alignment of spins within a triangular layer.
RMC runs with additional spins on the minor 2(d) sites did not improve the fit, so the minor spins are not decisive for the Q$_2$ feature.
\\
We summarise our PND results as follows: long-range correlations resulting in the Q$_1$=0.6 \AA$^{-1}$ peak arise mostly from the minor Fe-spins, while
 the short-range correlations at Q$_2$=0.77 \AA$^{-1}$ are magnetic in-plane correlations of the major Fe-spins.
To distinguish whether these are these static or dynamic we turned to inelastic neutron scattering, presented below.
%
\subsection{Magnetic dynamic correlations from inelastic neutron scattering}
\begin{figure}[ht]
\centering 
\includegraphics[width=0.495\columnwidth]{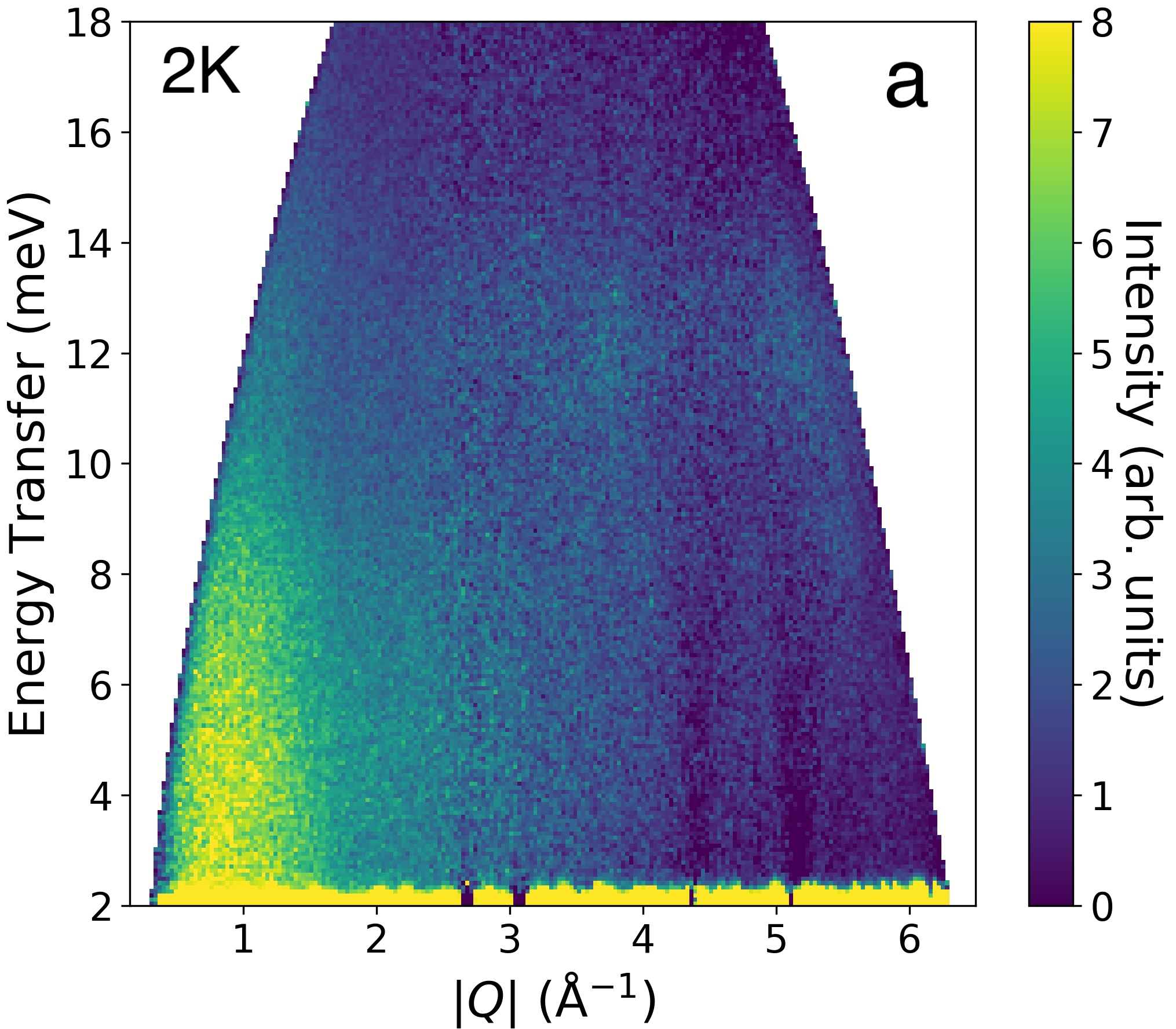}
\includegraphics[width=0.495\columnwidth]{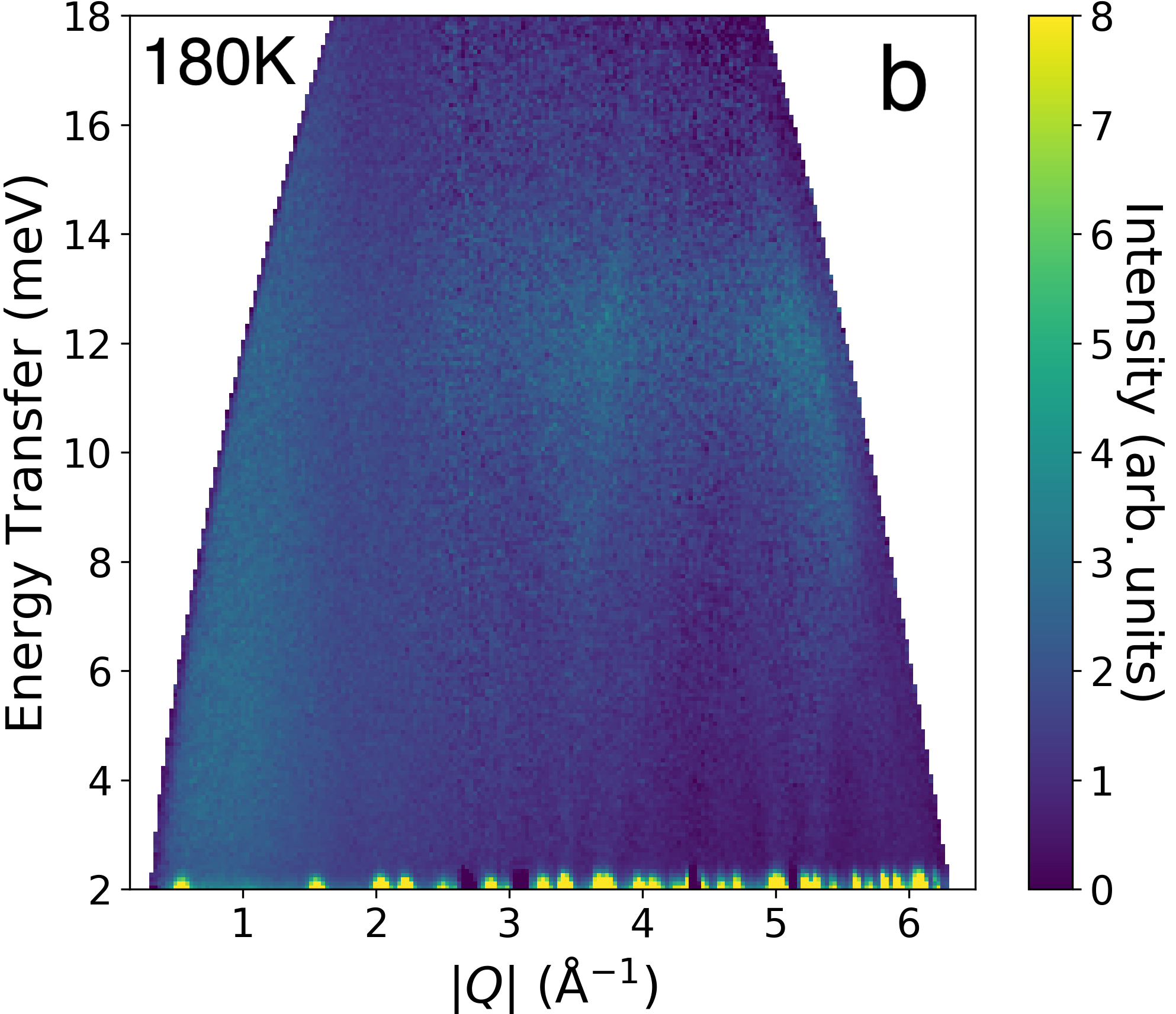}
\includegraphics[width=0.495\columnwidth]{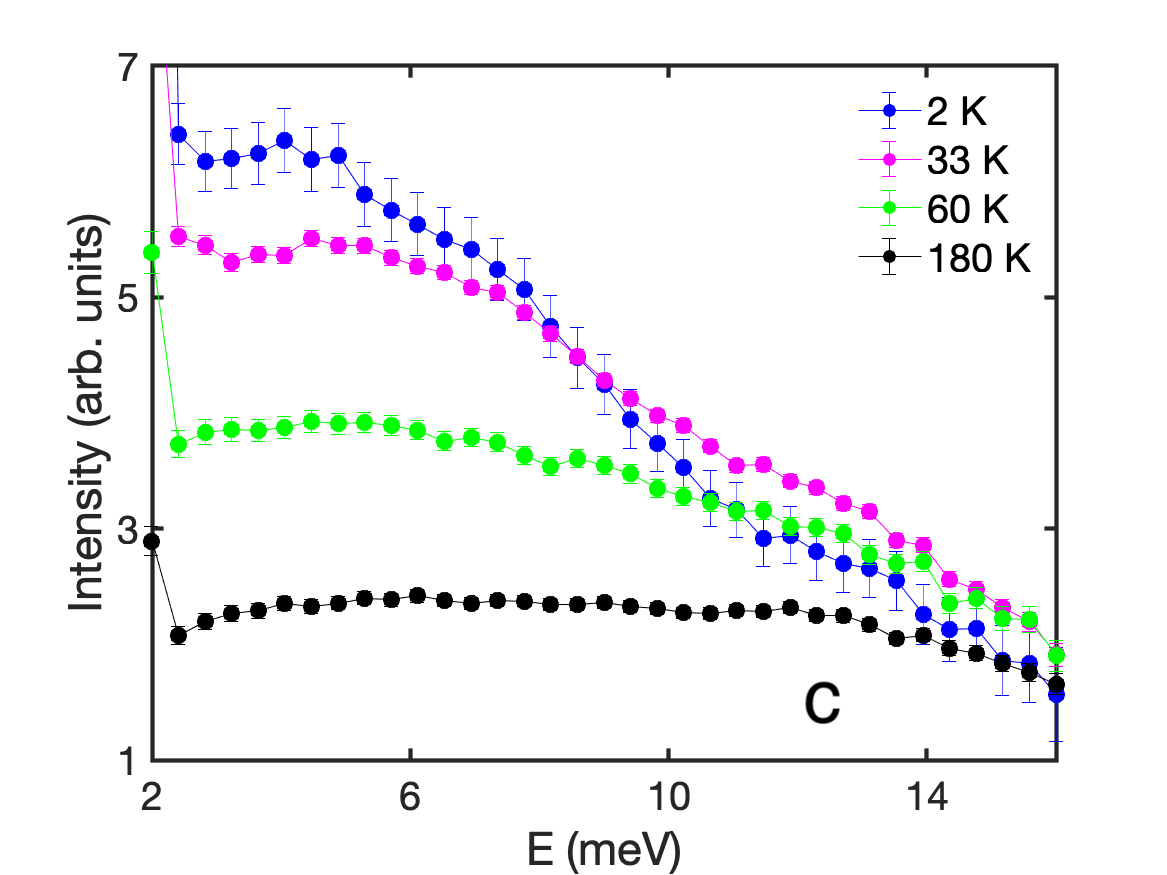}
\includegraphics[width=0.495\columnwidth]{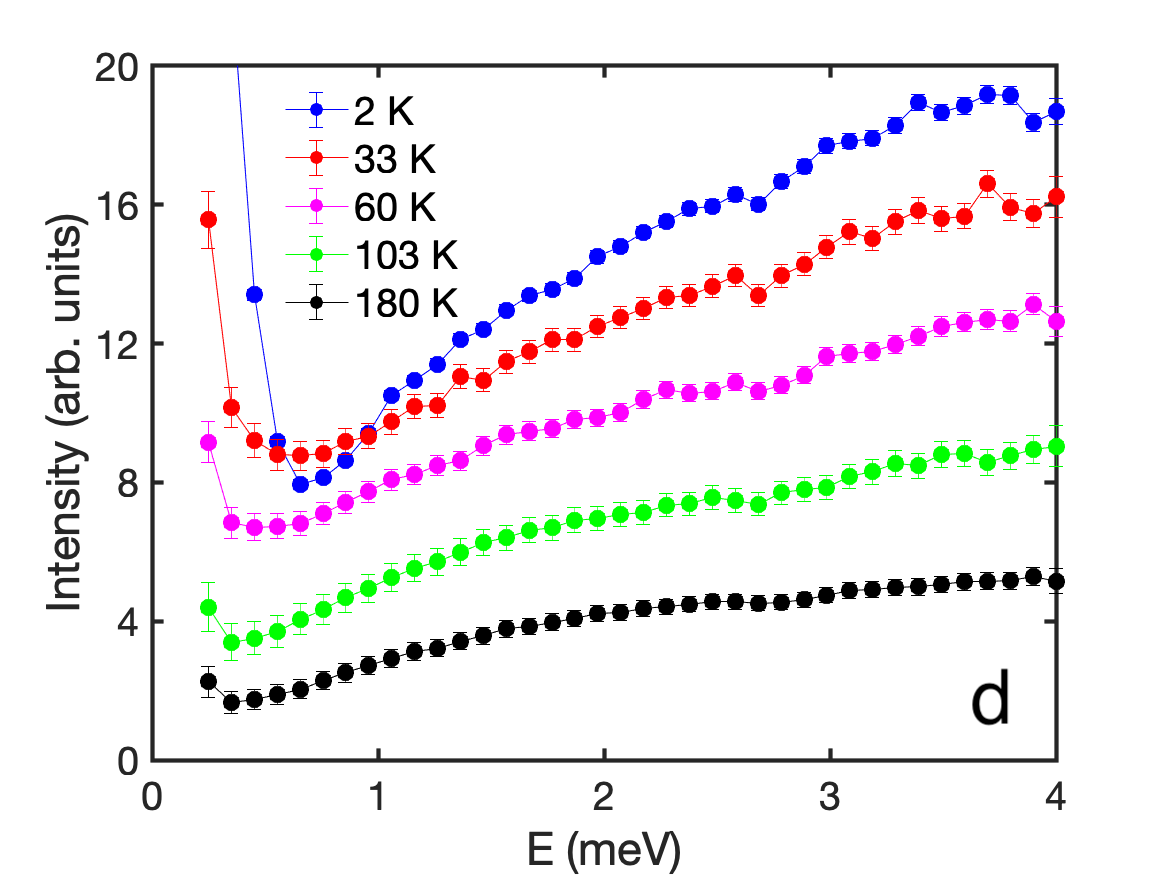}
\caption{Top: Excitation spectrum of {\FGS} measured on LET with $E_i$ = 24.3 meV at the lowest T= 2 K (a) and the highest 180 K (b) temperatures.
Bottom: Temperature evolution of the low-Q feature with signal integrated over Q=[0.3$\ldots$1.5] \AA$^{-1}$ with the $E_i$= 24.3 meV rep-mode ($\delta E_{0meV}$= 1.15 meV) (c) and over Q=[0.4$\ldots$0.8] \AA$^{-1}$, $E_i$ = 7.7 meV ($\delta E_{0meV}$ = 0.22 meV) (d). The Al can was subtracted and data were corrected for the Bose factor.}
\label{fig_let1} 
\end{figure}

\begin{figure}[ht]
\centering 
\includegraphics[width=0.30\columnwidth]{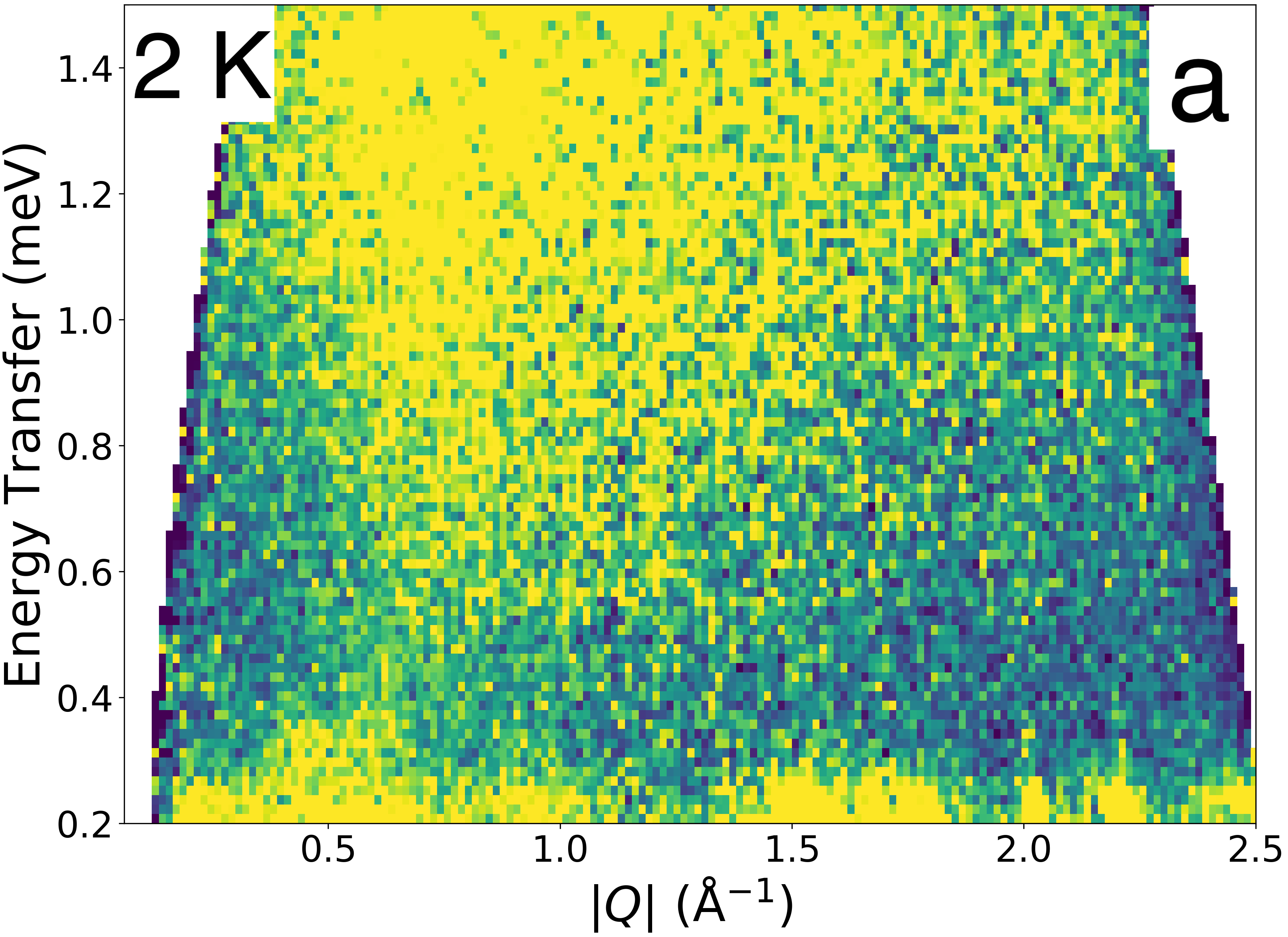}
\includegraphics[width=0.30\columnwidth]{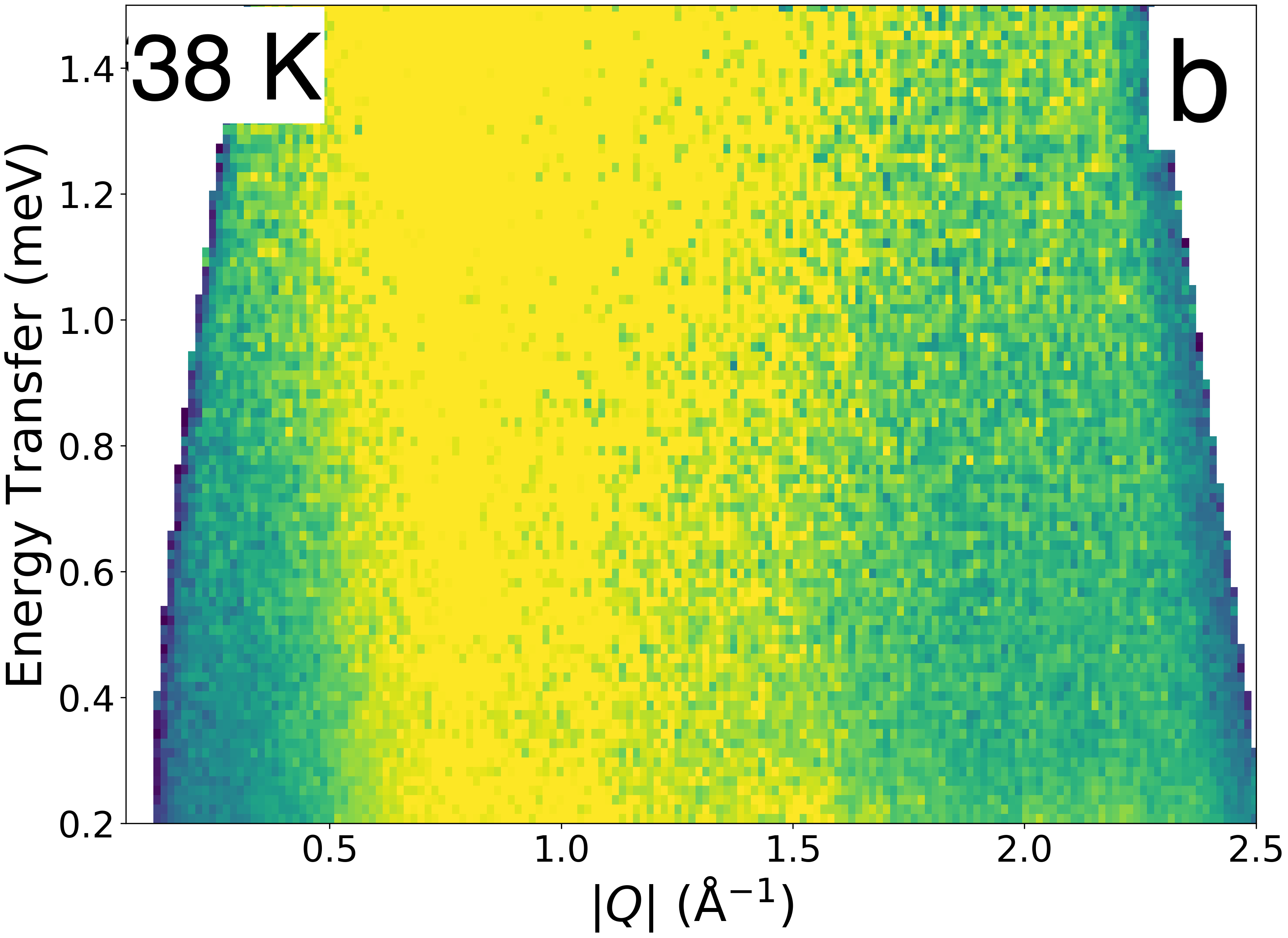}
\includegraphics[width=0.30\columnwidth]{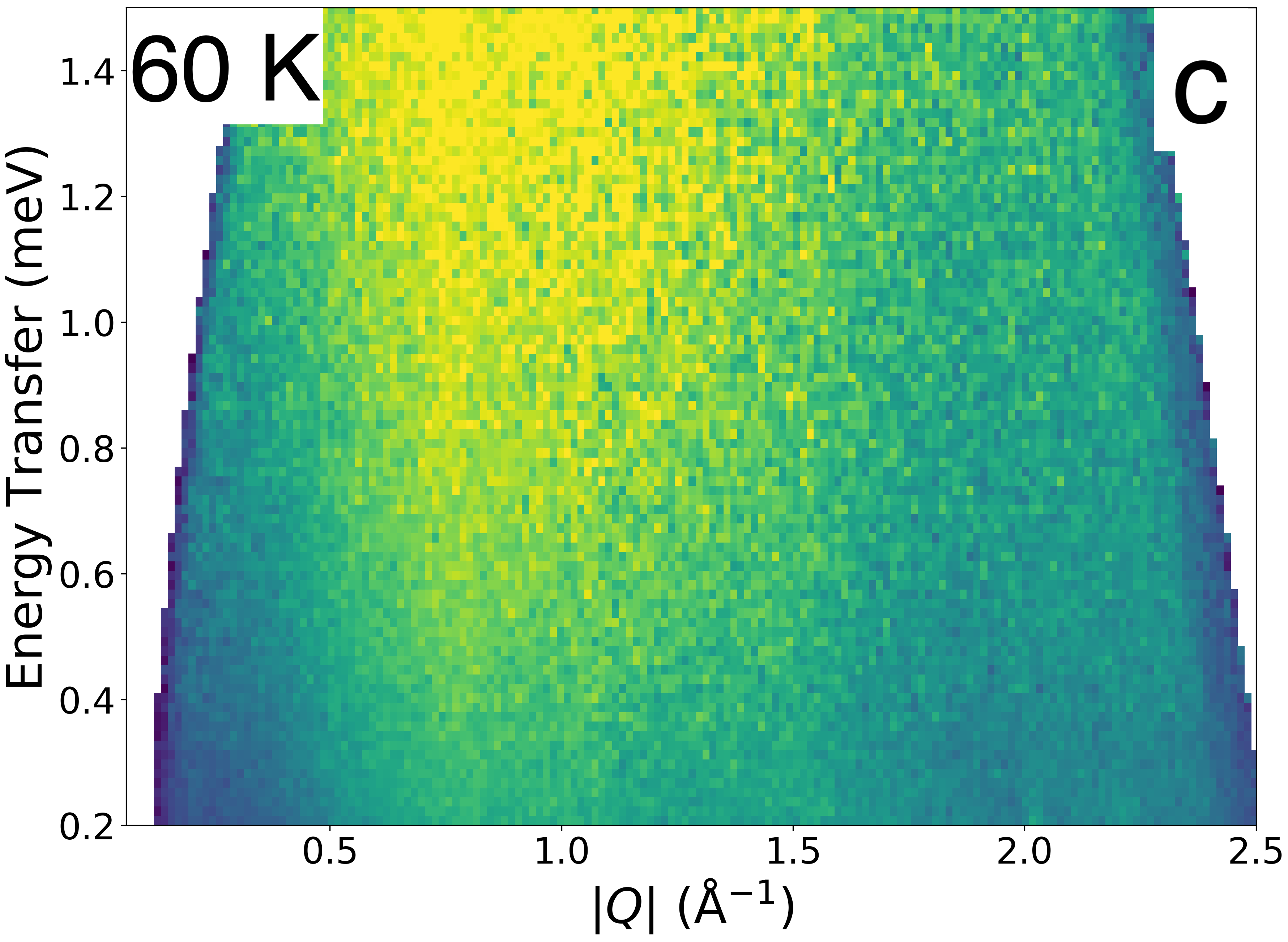}
\includegraphics[width=0.03\columnwidth]{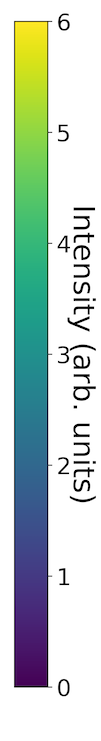}
\includegraphics[width=0.45\columnwidth]{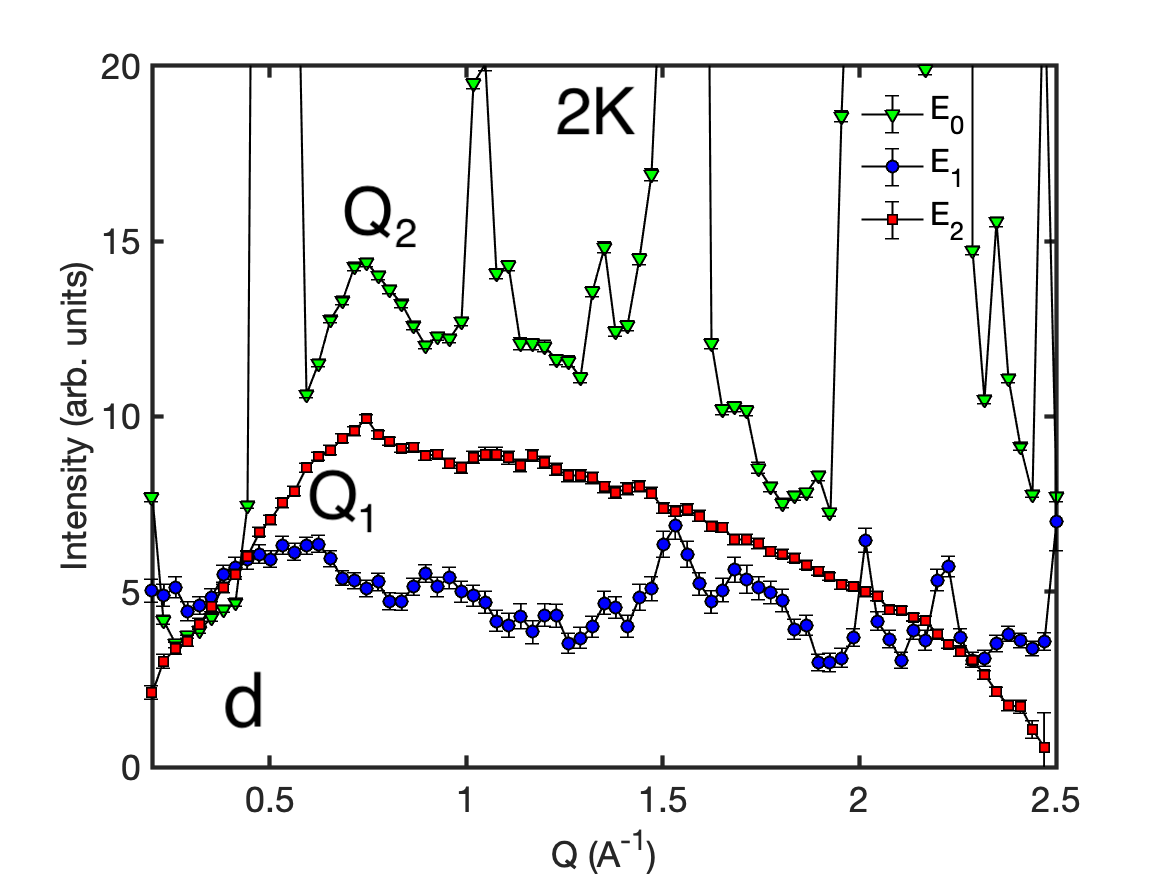} 
\includegraphics[width=0.45\columnwidth]{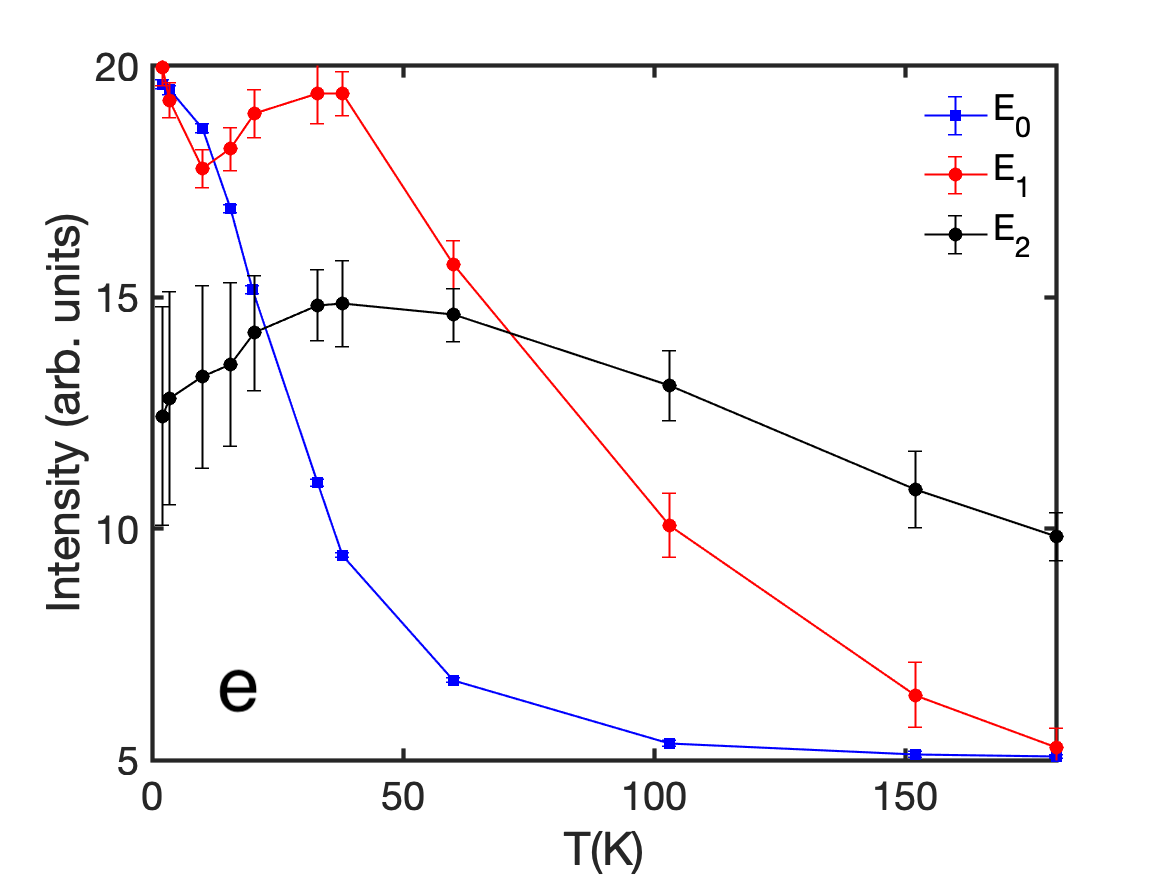}
\caption{Top: Part of the excitation spectrum of {\FGS} measured on LET with $E_i$= 3.8 meV ($\delta E_{0meV}$= 0.1 meV) at T= 2 K (a), 38 K (b), 60 K (c). 
Bottom: (d) 2 K elastic signal summed over E$_0$= [0$\ldots$0.2] meV (green symbols, signal divided by 50) and inelastic signals E$_1$= [0.2$\ldots$0.5] meV (blue symbols), E$_2$= [0.5$\ldots$3] meV (red symbols). (e) Temperature evolution of the elastic (blue squares) and inelastic (red and black circles) features. The elastic signal is integrated over Q=[0.6$\ldots$1] \AA$^{-1}$ and E$_0$=[-0.1$\ldots$0.1] meV, the inelastic signals are corrected for the Bose factor and integrated over Q=[0.44$\ldots$2] \AA$^{-1}$, E$_1$= [0.5$\ldots$2] meV and Q=[0.3$\ldots$1.5] \AA$^{-1}$, E$_2$= [11$\ldots$13] meV. The last curve is extracted from the $E_i$= 24.3 meV rep-mode.}
\label{fig_let2} 
\end{figure}

{\label{Sec3Sub3}}
%
The excitation spectrum of polycrystalline {\FGS} was measured on the LET spectrometer\cite{BewleyLET}.
Fig.~\ref{fig_let1} displays the wide-ranging S(Q, $\omega$) accessed with the $E_i$= 24.3 meV rep-mode at 2 K and 180 K. Two prominent INS features are evident. 
The first one is peaked at Q$_2$ = 0.77 \AA$^{-1}$ and E = 5 meV. It decreases towards the elastic line, suggesting that these might be gapped spin waves, but  it is hard to identify the value of the anticipated gap and temperature of its softening even from the lower energy $E_i$= 7.7 meV rep-mode spectrum (Fig.~\ref{fig_let1} d).
The second, high-energy feature at 13 meV spans over the whole measured interval of 1 $<$ Q $<$ 6 \AA$^{-1}$. The low Q-part (Q $<$ 4.5 \AA$^{-1}$) reduces with temperature, thus it has magnetic origin. The high Q part (Q $>$ Q=4.5 \AA$^{-1}$) persists, so it has significant phonon contribution. These two features reside on a very broad background which remains even at 180 K, the highest temperature reached in the experiment.\\
The anticipated gap region for the first feature is magnified in the top panel of Fig.~\ref{fig_let2}, where S(Q, $\omega$) maps measured with the $E_i$ = 3.8 meV rep-mode at three representative temperatures 2 K, 38 K and 60 K are presented. At 2 K the correlations are predominantly static. In Fig.~\ref{fig_let2}d the counts summed within 0 $<$ E$_1 <$ 0.2 meV are divided by a factor 50 to match the scale of the intensity within the gap 0.2 $<$ E$_2 <$ 0.5 meV and of the excitations above the gap 0.5 $<$ E$_2 <$ 3 meV. 
Interestingly, at 2 K a weak dispersive feature starting from Q$_1$ could also be identified. At higher temperatures it is completely obscured by a continuum centered at Q$_2$=0.77 \AA$^{-1}$. The continuum gains intensity in the 30-40 K region and is reduced with further increasing temperature.\\
It would be of interest to compare the temperature evolution of various static and dynamic correlations observed on LET and in bulk magnetic measurements. Unfortunately the insufficient resolution does not allow to separate the Q$_1$ feature from the nearby nuclear peak Fig.~\ref{fig_let2}d. Therefore we present the temperature dependence of the elastic intensity of the predominantly Q$_2$ feature summed within 0.6 $<$ Q $<$ 1 \AA$^{-1}$ in Fig.~\ref{fig_let2}e. It decreases sharply around 50 K. The two prominent INS signals centered at 5 and 13 meV, in contrast are maximal at 50 K. 
The 5 meV feature is integrated within the $E_i$= 3.8 meV setup and this allows the
documentation of the deflection point at 10 K (close to T$_f$).\\
The contribution of spin waves from Q$_1$=0.6 \AA$^{-1}$ is minute in the collected INS signal and cannot be disentangled from the broad major feature centered at Q$_2$ = 0.77 \AA$^{-1}$. The origin of the Q$_2$ feature evolves from static into dynamic with temperature increase and reaches maximum intensity at 50 K, which coincides with the broad maximum in the magnetic part of the specific heat.\cite{Nakatsuji2007} We presume it originates from the major spins.
In brief: The low energy part of INS data is dominated by the (5 meV, 0.77 \AA$^{-1}$) feature, the high energy part has a feature at 13 meV which is magnetic for Q $<$ 4.5 \AA$^{-1}$ and has a phonon contribution for Q $>$ 4.5 \AA$^{-1}$. The magnetic INS signal has maximum intensity at 50 K, where the elastic short-range correlations diminish. The broad scattering background beneath these features persist even at 180 K.

\section{Modelling of magnetic Hamiltonian}{\label{Sec4}}
%
We used several theoretical approaches to explain these variegated experimental observations. We start with the ideal triangular lattice and present analytical results for models with increasing number of bilinear exchanges and with the biquadratic exchange (Section. \ref{Sec4Sub1}). Then we test the model with three bilinear exchanges against our INS data (Section. \ref{Sec4Sub2}).
%
\subsection{Analytical results for models with bilinear and biquadratic exchanges}{\label{Sec4Sub1}}
%
First, we try to establish the range of exchange parameters which could give rise to magnetic correlations observed at Q$_1$=0.6 \AA$^{-1}$ for {\FGS} and 0.57 \AA$^{-1}$ for {\NGS}\cite{Nakatsuji2005} within the triangular lattice Heisenberg bilinear model. 
In this Section we use the theoretical notation (TN) for the reciprocal space (see Appendix {\ref{app:conv} for the definition of TN and its relation to the experimental notation (EN)).\\
We inspect analytically, with rising complexity, the single $J_{i=1,3}$, the  bilinear $J_{i=1,3}-J_{j=1,3}$ and the trilinear $J_1-J_2-J_3$ models on the two-dimensional triangular lattice described by the Hamiltonian:
\begin{equation}
H = J_1 \sum _{<i,j>}({\bf{S}}_i \cdot {\bf{S}}_j) +J_2 \sum _{<<i,j>>}({\bf{S}}_i \cdot {\bf{S}}_j) + J_3 \sum _{<<<i,j>>>}({\bf{S}}_i \cdot {\bf{S}}_j )
\label{eq:Hbi}
\end{equation}
Since we are interested in the region of the $J_1$, $J_2$, $J_3$ phase diagram with the incommensurately modulated ground states, we express the spin $\mathbf{S}_{i}$ on site $\mathbf{r}_{i}$ as\cite{Tamura2008}:
\begin{equation}
\mathbf{S}_{i}=\mathbf{n}_{1}\cos(\mathbf{k}\cdot\mathbf{r}_{i})-\mathbf{n}_{2}\sin(\mathbf{k}\cdot\mathbf{r}_{i}),
\label{eq:helix_d}
\end{equation}
where $\mathbf{n}_{1}$ and $\mathbf{n}_{2}$ are orthonormal vectors. 
For the Hamiltonian in Eq. \ref{eq:Hbi} we obtain the expression of the energy as a function of $\mathbf{k}$:
\begin{eqnarray}
E(\mathbf{k})&=&2 J_{1} N \left[\cos(k_x) + 2\cos(\frac{k_x}{2})\cos(\frac{\sqrt{3}k_y}{2})\right]+\\ \nonumber
&+&2 J_{2} N \left[2 \cos(\frac{3k_x}{2})\cos(\frac{\sqrt{3}k_y}{2})+\cos(\sqrt{3}k_y)\right]+\\ \nonumber
&+&2 J_{3} N \left[\cos(2k_x) + 2\cos(k_x)\cos(\sqrt{3}k_y)\right].
\label{eq:j1j2j3_d}
\end{eqnarray}
%
\begin{figure}[htb]
\includegraphics[width=0.49\columnwidth]{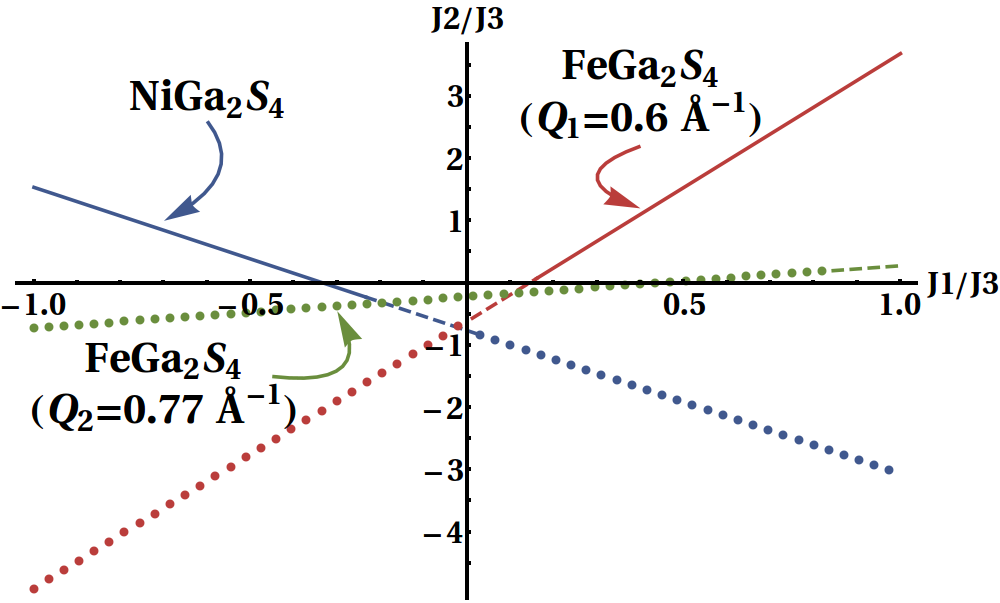}
\includegraphics[width=0.49\columnwidth]{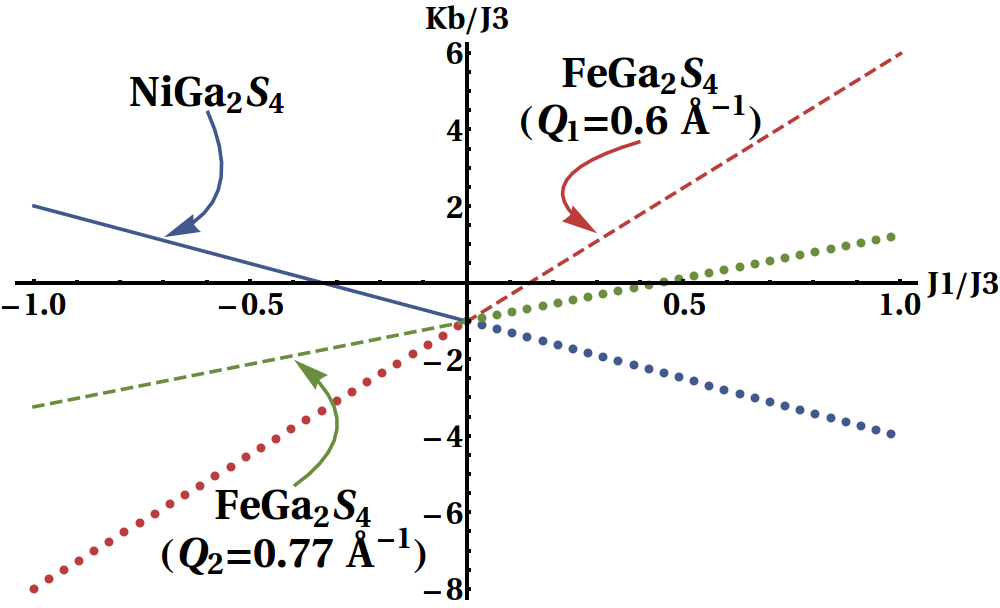}
\caption{The relation between (a) $J_{1}/J_{3}$ and $J_{2}/J_{3}$, and (b) $J_{1}/J_{3}$ and K$_{B}/J_{3}$ to obtain the ordering wave-vectors for {\NGS} (blue) and {\FGS} (red and green) within the $J_{1}-J_{2}-J_{3}$ and $J_{1}-J_{3}-K_{B}$ models. The critical points corresponding to a local maximum, a local minimum and a global minimum of the energy at the given values of the ordering wave-vectors are distinguished by the dotted, dashed and solid lines, respectively.}
\label{fig:j1j2j3}
\end{figure}
%
It is important to note that there is no explicit dependence of the energy on the vectors $\mathbf{n}_{1}$ and $\mathbf{n}_{2}$, reflecting the SO(3) global symmetry of the model. In order to find the possible ordering vectors we minimize $E(\mathbf{k})$ with respect of $\mathbf{k}$.
Depending on the values of the parameters $J_{1}$, $J_{2}$ and $J_{3}$, the minima of this model lie along the following two sets of directions: (i): $\pm(1,0)$, $\pm(\frac{1}{2},\frac{\sqrt{3}}{2})$, $\pm(-\frac{1}{2},\frac{\sqrt{3}}{2})$, and (ii): $\pm(\frac{\sqrt{3}}{2},\frac{1}{2})$, $\pm(0,1)$ and $\pm(\frac{\sqrt{3}}{2},\frac{1}{2})$.
Since the ordering vector $\bf{k}\approx$({\six} {\six} 0) in EN observed in {\MGS} (M= Ni, Fe) corresponds to the set (i), we focus on this direction and search for a minimum along the $\mathbf{\hat{x}}$ direction (or equivalently along any direction of this set). With this condition the value of $k$ should satisfy:
\begin{equation}
\frac{J_{1}}{J_{3}}=-\frac{2(\sin(k) + \sin(2k))+3 \frac{J_{2}}{J_{3}} \sin(\frac{3k}{2})}{\sin(\frac{k}{2}) + \sin(k)}.
\label{eq:j1j2j3min}
\end{equation}
From this equation we obtain linear relations between $J_1$, $J_2$ and $J_3$ presented in Fig.~\ref{fig:j1j2j3} a. 
This linear relation expresses the values of the parameters for which the energy develops a critical point. The critical points gain stability when we move along
each line, running from a local maximum (represented by dots) to a local minimum (dashed segment) and finally to a global minimum (solid line). For {\FGS} the Q$_2$ vector (green line) becomes a global minimum for $J_1/J_3 >$ 1.72. Further restrictions on $J_{1}/J_{3}$ should be included for these critical points to belong in the set (i): for  {\NGS} $J_1/J_3 >$ -1 and to obtain either the Q$_1$ or Q$_2$ vectors for {\FGS} requires $J_{1}/J_{3} <$ 0.6 and $J_{1}/J_{3} <$ 4.3, respectively.

The sign of the slope for a modulated state with $k\approx 0.155\times4\pi$ for {\NGS} (blue) and with $k\approx 0.1737\times4\pi$ for {\FGS} (red) is different. This change of the sign takes place at the critical wave vector $\bf{k}$=({\six} {\six} 0).\\
It is important to note that the single $J_{i=1,3}$ - exchange models and the bilinear $J_1-J_2$ and $J_1-J_3$ exchange models are not sufficient to explain the Q$_1$=0.6 \AA$^{-1}$ feature observed for {\FGS}. For example, the ordering vector $k\approx 0.155\times4\pi$ for {\NGS} can be obtained from the $J_1-J_3$ model\cite{Tamura2011}, this is not the case for $k\approx 0.1737\times4\pi$ for {\FGS}. In those models the ordering vector implies either a uniform (ferromagnetic) state, a commensurate 120 $ \deg$ structure or it does not match the direction (the set indicated in (i)) signalled by the experimental results. For the $J_1-J_2-J_3$ exchange model such solutions, however, are available.\\
Now we consider the ordering vector with the length $k\approx 0.222\times4\pi$, which could correspond to the center of the diffuse feature at Q$_2$=0.77 \AA$^{-1}$ observed for {\FGS}. 
The parameters of the ideal triangular lattice model are related through a linear equation depicted in Fig.~\ref{fig:j1j2j3} by the green line. These solutions strongly differ from the Q$_1$ lines and require significant $J_1$ coupling.\\
We proceed with the 'model engineering' and raise the complexity by adding the biquadratic interaction $K_B$ between first nearest neighbors $K_B \sum _{<i,j>}({\bf{S}}_i \cdot {\bf{S}}_j)^2$. 
We repeat the previous procedures for two models: $J_1-J_3-K_B$ and $J_{1}-J_{2}-J_{3}-K_{B}$.\\
The energy of a modulated state for the $J_1-J_3-K_B$ model reads:
\begin{eqnarray}
E(\mathbf{k})&=&2 J_{1} N \left[\cos(k_x) + 2\cos(\frac{k_x}{2})\cos(\frac{\sqrt{3}k_y}{2})\right]+\\ \nonumber
&+&2 J_{3} N \left[\cos(2k_x) + 2\cos(k_x)\cos(\sqrt{3}k_y)\right]+\\ \nonumber
&+&2 K_{B} N \left[3+\cos(2k_x) + 2\cos(k_x)\cos(\sqrt{3}k_y)\right],
\label{eq:j1j3kb}
\end{eqnarray}
As we did before we look for the minimum of the energy along the $\mathbf{\hat{x}}$ direction. With this condition the value of $k$ should satisfy:
\begin{equation}
\frac{J_{1}}{J_{3}}=-\frac{2(\frac{K_{B}}{J_{3}}+1) (\sin(k) + \sin(2k))}{\sin(\frac{k}{2}) + \sin(k)}.
\label{eq:j1j2kbmin}
\end{equation}
From this equation we obtain the relations between the values of $J_1$, $J_3$ and $K_B$  (Fig.~\ref{fig:j1j2j3}) right for which modulated states with $k\approx 0.155\times 4\pi$ for {\NGS}, $k\approx 0.1737\times 4\pi$  and $k\approx 0.222\times 4\pi$ for {\FGS} are the ground states. The role of K$_B$ is significant, but in the real system we expect K$_B$ to be a perturbative and not the leading parameter.

Finally, if we add the $J_{2}$ exchange interaction to the previous model, the energy as a function of $\mathbf{k}$ is:
\begin{eqnarray}
E(\mathbf{k})&=&2 J_{1} N \left[\cos(k_x) + 2\cos(\frac{k_x}{2})\cos(\frac{\sqrt{3}k_y}{2})\right]+\\ \nonumber
&+&2 J_{2} N \left[2 \cos(\frac{3k_x}{2})\cos(\frac{\sqrt{3}k_y}{2})+\cos(\sqrt{3}k_y)\right]+\\ \nonumber
&+&2 J_{3} N \left[\cos(2k_x) + 2\cos(k_x)\cos(\sqrt{3}k_y)\right]+\\ \nonumber
&+&2 K_{B} N \left[3+\cos(2k_x) + 2\cos(k_x)\cos(\sqrt{3}k_y)\right].
\label{eq:encomp}
\end{eqnarray}
Since we are interested in the states with $\mathbf{k}$ along the $\mathbf{\hat{x}}$ direction, we look for the minimum of the energy along this direction. With this condition the value of $k$ should satisfy:
\begin{equation}
\frac{J_{1}}{J_{3}}=-\frac{2(\frac{K_{B}}{J_{3}}+1)(\sin(k) + \sin(2k))+ 3 \frac{J_{2}}{J_{3}} \sin(\frac{3 k}{2})}{\sin(\frac{k}{2}) + \sin(k)}.
\label{eq:j1j2j3kbmin}
\end{equation}
From this linear equation the values of $J_1$, $J_2$ and $K_B$ for which the ICM ground states with $k\approx 0.155\times 4\pi$, $k\approx 0.1737\times 4\pi$ and $k\approx 0.222\times 4\pi$ are obtained.\\
Our analytical approach shows that the parameter space of the ideal triangular model which fits the experimental ordering wave vectors is quite extended, however constrained by Eq.\ref{eq:j1j2j3kbmin}, and additional input is necessary to select the unique set of exchange parameters. Despite the broad range of parameters, we emphasize on the relevance of the constraints imposed by Eqs.\ref{eq:j1j2j3min},  \ref{eq:j1j2kbmin} and \ref{eq:j1j2j3kbmin} for searching the correct set of couplings.
%
\subsection{Extraction of exchange parameters from the INS spectrum using linear spin wave theory}{\label{Sec4Sub2}}
%
\begin{figure}[ht]
\includegraphics[width=\textwidth]{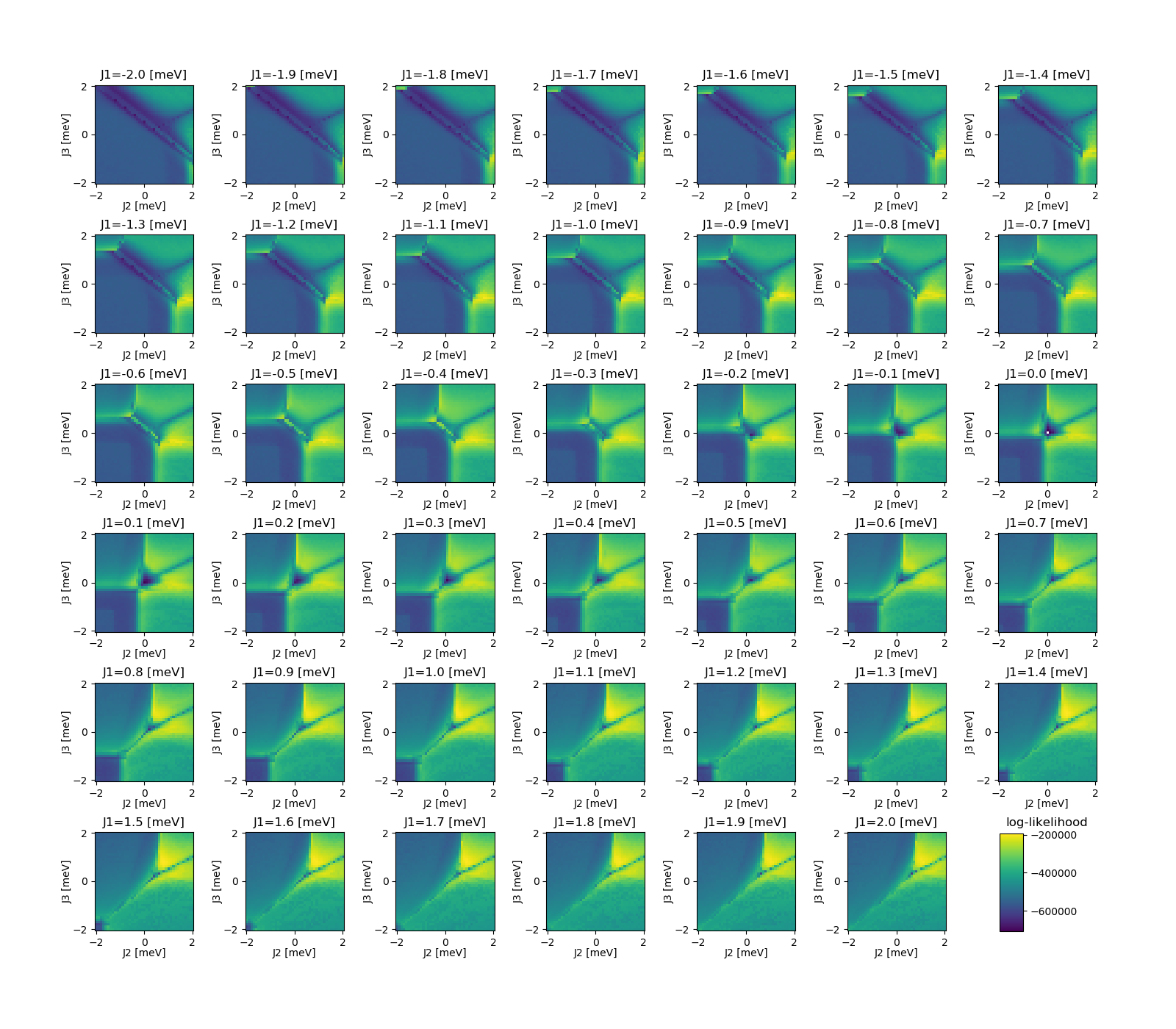}
\caption{Maximum log-likelihood after optimising the magnitude of the spin wave contributions.}
\label{fig:llh}
\end{figure}
\begin{figure}[ht]
  \includegraphics[width=0.49\textwidth]{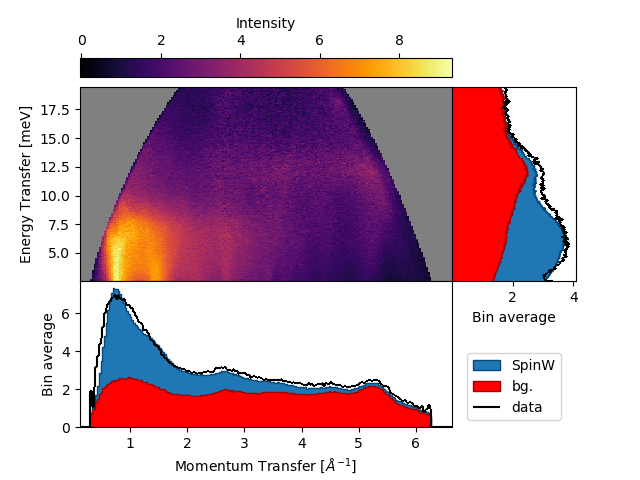}
  \includegraphics[width=0.49\textwidth]{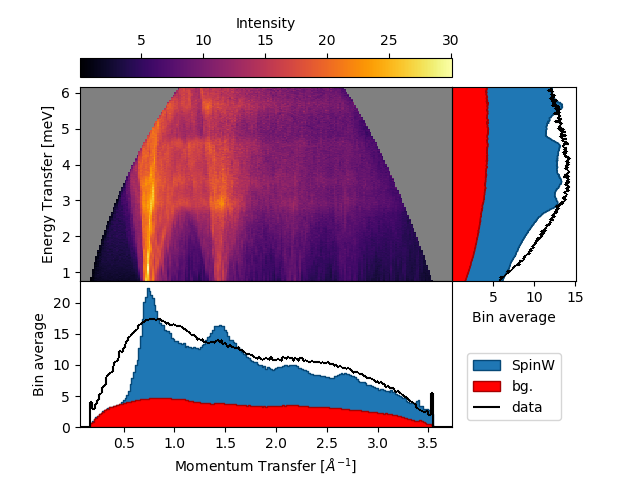}
  \caption{
Predicted $E_i=24.3$\,meV (left) and $E_i=7.7$\,meV (right) INS spectra at 2\,K, constructed from an empirical background model measured at 180\,K and spin wave simulations for the set of the exchange values $J_1=1.7$\,meV, $J_2=0.9$\,meV, $J_3=0.8$\,meV.
The margin plots show comparisons with data measured at 2\,K.
  }
  \label{fig:plot_15_08_08}
\end{figure}
%
We narrow the exchange parameter space calculating the spin excitation spectrum and fitting it to the experimental S(Q, $\omega$). 
We consider the triangular lattice Heisenberg bilinear model of Eq.\ref{eq:Hbi} and focus on the Q$_2$=0.77\,\AA$^{-1}$ feature observed in the INS spectra assuming that it emerges from the spin waves.
We take the 180\,K data after the Bose factor correction as an empirical background (even though it contains a featureless magnetic contribution) and attribute the rest of the 2\,K data to spin wave excitations.
Using the ground state of Eq.\ref{eq:j1j2j3_d}, we calculate spin wave excitations with the SpinW program\cite{SpinW} and compare the resulting spectra with the INS data measured at 2\,K.
To avoid regions in data that are saturated by elastic scattering, only energy transfers above 2.4\,meV and 0.75\,meV are considered for the two setups of $E_i=24.3$\,meV and $E_i=7.7$\,meV, respectively.
In the simulation, the experimental energy resolution of each setup is approximated by a constant corresponding to its nominal (energy-dependent) value at the peak of the magnetic excitation.
No intrinsic broadening is considered.
The nominal uncertainties on the measured intensity are obtained from the data itself and are assumed to be underestimated for bins with low neutron counts.
Therefore, these uncertainties are thresholded to a minimum value determined by the actual variance across low-intensity bins in regions of data without large scale features.
The model is evaluated on a discretised grid of values ranging from -2\,meV to 2\,meV in increments of 0.1\,meV for each of exchange parameters $J_1$, $J_2$, $J_3$.
Figure~\ref{fig:llh} shows the maximum log-likelihood at each point after optimising the magnitude of the spin wave contributions.
The maximum log-likelihood suggests that $J_1$, $J_2$ and $J_3$ are antiferromagnetic.
The best fit is obtained for $J_1=1.7$\,meV, $J_2=0.9$\,meV, $J_3=0.8$\,meV, which is consistent with the relation imposed by Eq. \ref{eq:j1j2j3min}.\\
We note that the general locations of the observed excitations can be approximately reproduced by this model, however the detailed structure of the simulated spin waves is not evident in the data, and this set of parameters is out the range indicated in Figure~\ref{fig:j1j2j3} a. In particular the $E_i=7.7$\,meV setup reveals that the predicted drop in intensity at low energy transfer is less rapid than what can be observed in data. These discrepancies are not removed by inclusion of the single-ion anisotropy $D_n \sum_i (S_i)^2$, which should be 20\% of exchange and planar according to the magnetic susceptibility measurements.\\
We conclude that the best fit to our powder INS data provides the set of exchange parameters within the $J_{1}-J_{2}-J_{3}$ bilinear model but this model is not sufficient to explain the whole observed excitation spectrum.
%
\section{Summary and Discussion}{\label{Sec5}}
%
Our extensive experimental study of the triangular antiferromagnet {\FGS} implies that this material, and most probably also {\NGS}, has significant mixing of the M  (M = Ni, Fe) and Ga cations between the octahedrally coordinated 1(b) and tetrahedrally coordinated 2(d) positions. Nearly 20\% of the 1(b) sites forming magnetic triangular layers are occupied by nonmagnetic ions and nearly 10\% of the 2(d) sites forming non-magnetic triangular layers host magnetic M-ions.

This mixing is an important ingredient of highly nontrivial magnetic properties. It introduces a long-rage magnetic order at T$_N$=5.5 K revealed by the Q$_1$=0.6 \AA$^{-1}$ peak in neutron powder diffraction, which we associated with the $\bf{k}$=(0.1737(1) 0.1737(1) 0) propagation vector. This magnetic order is dominated by the amplitude modulated $z$-component of the minor Fe-spins at the 2(d) sites. The amount of the minor Fe-ions are well below the percolation limit of the stacked triangular antiferromagnetic lattice (p$_c^{side}$=0.26240(5)\cite{Schrenk2014}), so the layers of the major Fe ions possibly mediate this exchange. 

The broad feature at Q$_2$=0.77 \AA$^{-1}$  in neutron powder diffraction corresponds to short-range magnetic correlations within the triangular layers of the major Fe sites. These in-plane correlations are static at low temperatures, evolve into dynamic correlations at elevated temperatures 30 - 50 K and fade away at temperatures comparable to $|\Theta|_{\rm CW}$=160 K. These dynamic correlations could be confined spin waves, the so-called Halperin-Saslow modes\cite{Halperin1977, Podolsky2009} of the major Fe moments, which freeze in due to numerous nonmagnetic impurities. The nonmagnetic defects possibly trigger formation-dissociation of the Z$_2$ vortices. According to the MC simulations\cite{Ajiro1988} the characteristic temperature T$_{KM}$ related to the pairing-dissociation of the Z$_2$ vortices is correlated with the effective exchange as
T$_{KM}$ =0.66 $J$ S$^2$. For our case $J$ = 13 K (estimated from susceptibility) and S=2, which this results in T$_{KM}$ = 34 K.
This temperature corresponds to the anomalies found in muon spectroscopy (T*= 33 K) \cite{Reotier2012, Zhao2012} and to the change between  static and dynamic nature of magnetic soft range correlations in our neutron scattering. Thus, even if the {\MGS} family is not an ideal  TLAFM system, it might realize interesting phenomena predicted theoretically for this model.\\
Our extensive theoretical study of the ideal triangular lattice bilinear-biquadratic Heisenberg model has value by itself. The analytical study provided in Section {\ref{Sec4Sub1}} could be applied to a generic triangular system and as such serves as guide to study other compounds. It rationalizes the ratios $J_1/J_3$, $J_2/J_3$ and $K_B/J_3$ for a provided ground state propagation vector. For the specific case of {\FGS} it notifies that the Q$_1$ and Q$_2$ features require possibly different sets of exchange parameters as they reside on the lines with one intercept point.
We reduce the range of possible exchange parameters for {\FGS} by modelling the distinct features of the experimental low-temperature INS spectra as conventional spin waves within the $J_1-J_2-J_3$ bilinear Hamiltonian. The best fit is obtained for the set of AF couplings $J_1$=1.7\,meV, $J_2$=0.9\,meV, $J_3$=0.8\,meV, which is close to the analytical solutions for Q$_2$. However, only the gross features of the INS spectra are explained. The origin of the broad scattering background remains elusive and for its understanding it is important to disentangle magnetic properties of inversion-free and inversion-containing triangular antiferromagnetic materials.\\
Our results imply that the short-range correlations at Q$_2$ and the anomalies around 30 K - 50 K are the properties of the major magnetic sites of the {\FGS} system. The main exchange couplings are antiferromagnetic and $J_1 > J_3 \approx J_2$. How strongly the behaviour of major Fe triangular sublattice is influenced by the minor Fe sites remains unclear. 
Theoretical predictions are presently limited to only small amount of impurities and to the 
first nearest-neighbors\cite{Wollny2011, Maryasin2013}. Extensions to the experimentally found values can bring answers to this question. 
Preliminary Monte Carlo simulations with the experimentally observed amount of impurity seem to reproduce
the Q$_1$ and Q$_2$ features observed in inelastic neutron scattering. However further exploration would be necessary. Novel synthetic routes to obtain inversion free members of the {\MGS} family could in the future facilitate experimental validation of the ideal TLAFM model.\\
%
\begin{acknowledgments}
This work was performed at SINQ, Paul Scherrer Institute, Villigen, Switzerland with financial support of the Swiss National Science Foundation (Grant No. 200020-182536) and Centro Lanitoamericano-Suizo (Seed money Grant No. SMG1811) and the Swiss Data Science Center (Project BISTOM C17-12).
We acknowledge F. Krumeich for recording SEM images and EDXS analysis.
\end{acknowledgments}
%
\section{Appendix A: Materials and methods}{\label{Sec6}}
%
Polycrystalline {\FGS} and {\NGS} samples were prepared by solid state synthesis from FeS (NiS) and GaS starting materials. The mixture was annealed for one week at 900 deg C  in an evacuated quartz ampule. Then the sample was homogenised by grinding, resealed, annealed again and cooled within 24 hours. Single crystals were grown by the chemical vapor transport method using iodine as a transport agent and a temperature gradient 1000/900 deg C. The crystals were then quenched.

The purity of the samples and the details of the crystal structure for {\FGS} in the temperature range 5 K - 300 K were determined for polycrystalline material on the x-ray powder diffractometer of the X04SA Material Science beamline at the SLS synchrotron ($\lambda$=0.641090 \AA) and for {\FGS} and {\NGS} single crystals on a Bruker SMART diffractometer equipped with an Apex I detector and an Apex II detector (Mo K$\alpha$=0.71073 \AA) at ETH Z\"{u}rich at 300 and 100 K, respectively.
CSD 2078453 and CSD 2078454 contain the supplementary crystallographic data for {\FGS} and {\NGS}, respectively. These data can be obtained free of charge from FIZ Karlsruhe via  www.ccdc.cam.ac.uk/structures.
Powder patterns were refined using the Fullprof suite\cite{fullprof1993}, for single crystal refinements Shelx\cite{shelx} was used. Scanning electron microscopy (SEM) images of the as-obtained samples were recorded with a secondary electron detector on a Gemini 1530 (Zeiss) microscope. Energy-dispersive X-ray spectroscopy (EDXS) was performed with a SSD system (Noran) attached to this microscope.

Magnetic susceptibility was measured in the temperature range 1.8 K - 350 K in an applied magnetic field of 0.1 T using a Magnetic Properties Measurement System superconducting quantum interference device (SQUID) magnetometer, Quantum Design. 

Neutron powder diffraction (NPD) patterns for {\FGS} were measured on the DMC diffractometer at SINQ ($\lambda$=2.4575 \AA).
Data were collected at several temperatures between 1.3 K and 180 K with the measuring time up to 12 h per pattern. 
XYZ-polarized neutron powder diffuse scattering was measured on the  D7 spectrometer at ILL (E$_i$=3.8 meV) at 1.4 K (4 h) and 60 K (10 h). 
The magnetic scattering was extracted from  isotropic magnetic scattering the magnetic ${\DSC}_{mag}$, incoherent ${\DSC}_{inc}$ and nuclear ${\DSC}_{nuc}$ cross sections were evaluated by the following equations\cite{Stewart2008}:
\begin{equation}
{\DSC}_{mag} = 2{\DSC}^x_{sf}+2{\DSC}^y_{sf}-4{\DSC}^z_{sf}
\end{equation}
where $x, y, z$ refer to the direction of the incident polarization, $sf$ and $nsf$ stands for spin-flip and non-spin-flip. 

Time-of-flight inelastic neutron scattering (INS) was measured for {\FGS} on the LET spectrometer\cite{BewleyLET} at ISIS. We used
the Rep-Rate Multiplication mode with the Fermi chopper rotating at 240 Hz in the high-flux setup, which allowed in a single time frame to collect data from incident neutron energies of 1.5, 2.24, 3.38, 7.7 and 24.3 meV. The Al can was subtracted and data were corrected for the Bose factor.
\section{Appendix B: Conventions and notations for the reciprocal space}{\label{app:conv}}

Two different notations can be encountered in the literature depending on the type of article (experimental or pure theoretical). We distinguish them by the names ``experimental notation'' (EN) and ``theoretical notation'' (TN). 

In the EN notation a vector $\mathbf{k}$ in the reciprocal space is expressed in terms of the basis vectors $\mathbf{b}_{1}$ and $\mathbf{b}_{2}$ as follows $\mathbf{k}=k_{1}\mathbf{b}_{1}+k_{2}\mathbf{b}_{2}$.
In the TN notation a vector $\mathbf{k}$ is expressed in terms of the basis vectors $\hat{\mathbf{x}}$ and $\hat{\mathbf{y}}$ as follows $\mathbf{k}=k_{x}\hat{\mathbf{x}}+k_{y}\hat{\mathbf{y}}$.
\begin{figure}[htb]
\includegraphics[width=0.99\columnwidth]{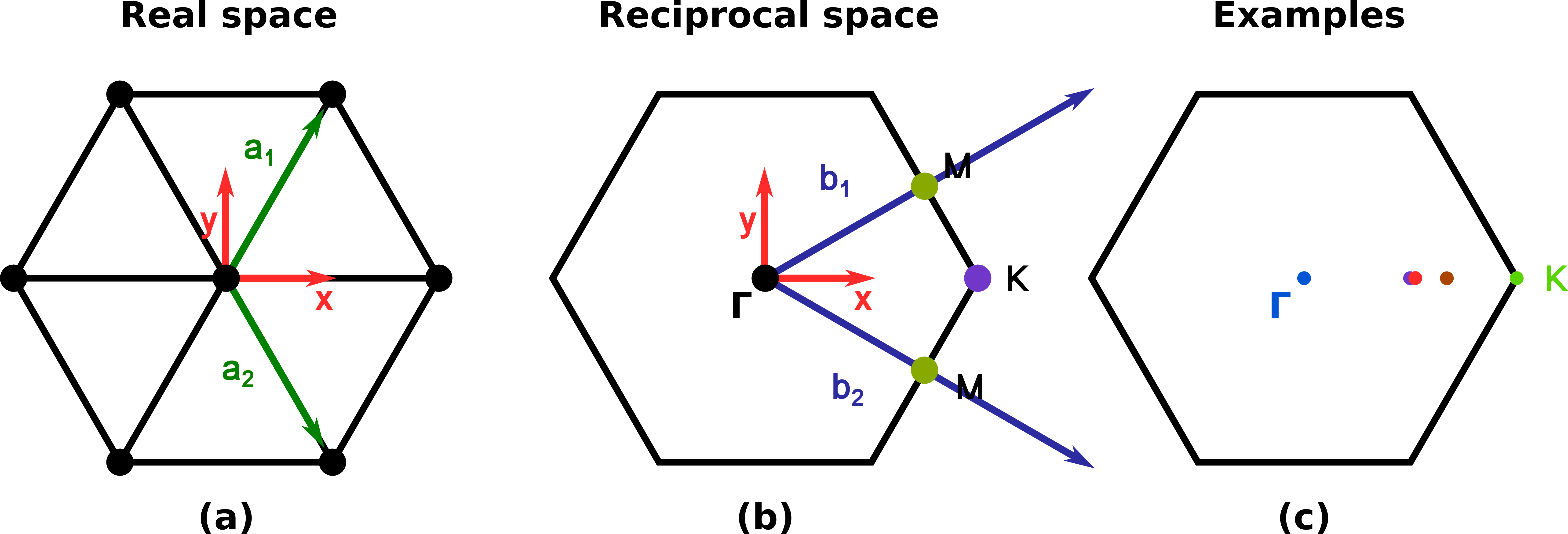}
\caption{
The red arrows represent the $\hat{\mathbf{x}}$ and $\hat{\mathbf{y}}$ unit vectors in all the cases. (a) Basis vectors $\{\mathbf{a}_{1},\mathbf{a}_{2}\}$ (in green) for the real space of the triangular lattice. (b) The first Brillouin zone for the triangular lattice. The M point (green circles), K point (purple circle) and $\Gamma$ point (black circle). The blue arrows represent the basis vectors $\mathbf{b}_{1}$ and $\mathbf{b}_{2}$ in the reciprocal space. (c) Some examples of interest, from the $\Gamma$ point (0, 0, 0), in light-blue, to the K point ({\third},{\third},0), in light-green, we have the points (expressed in CN): ({\six},{\six},0), in purple, (0.1737, 0.1737,0), in red, and (0.222, 0.222,0), in brown.}
\label{fig:fbx_convenciones}
\end{figure}
The connection between both notations is given by a change of basis (from $\{\mathbf{b}_{1},\mathbf{b}_{2}\}$ to $\{\hat{\mathbf{x}},\hat{\mathbf{y}}\}$).
The basis vectors $\mathbf{b}_{1}$ and $\mathbf{b}_{2}$ (in blue) are expressed in terms of the vectors $\hat{\mathbf{x}}$ and $\hat{\mathbf{y}}$  (in red) as follows:
\begin{eqnarray}
\mathbf{b}_{1}&=&\frac{2\pi}{a}\hat{\mathbf{x}}+\frac{2\pi}{\sqrt{3}a}\hat{\mathbf{y}},\\
\mathbf{b}_{2}&=&\frac{2\pi}{a}\hat{\mathbf{x}}-\frac{2\pi}{\sqrt{3}a}\hat{\mathbf{y}}.
\end{eqnarray}

So, for instance, in EN the state $({\third}, {\third},0)$ (the K-point) corresponds to the triplet $(\frac{4\pi}{3a},0,0)$ in TN,
\begin{equation}
\frac{1}{3}\mathbf{b}_{1}+\frac{1}{3}\mathbf{b}_{2}=\frac{4\pi}{3}\hat{\mathbf{x}}.
\end{equation}

The M-point in EN is identified with (\{\half}, 0, 0), and in the TN with the triplet $(\frac{\pi}{a},\frac{\pi}{\sqrt{3}a},0)$, 
\begin{equation}
\frac{1}{2}\mathbf{b}_{1}=\frac{\pi}{a}\hat{\mathbf{x}}+\frac{\pi}{\sqrt{3}a}\hat{\mathbf{y}}.
\end{equation}

Thus, when in the EN notation we refer to the $(0.155, 0.155)$ wave vector, it corresponds to the wave vector $k\approx 0.155\times 4\pi$ in the TN notation (taking $a=1$).
 %

%

\begin{thebibliography}{99}
\bibitem{Anderson1973} P. W. Anderson, Mat. Res. Bull  {\bf8}, 153 (1973).
\bibitem{Jolicoeur1990} Th. Jolicoeur, E. Dagotto, R. Gagliano, S. Bacci, Phys. Rev. B  {\bf42}, 4800(R) (1990).
\bibitem{Okubo2012} T. Okubo, S. Chung, and H. Kawamura, Phys. Rev. Lett. {\bf108}, 017206 (2012).
\bibitem{Dogguy1980} L. D. Smiri, N. H. Dung, and M. P. Pardo,  Materials Research Bulletin {\bf15} 861 (1980).
\bibitem{Nakatsuji2007} S. Nakatsuji, H. Tonomura, K. Onuma, Y. Nambu, O. Sakai, Y. Maeno, R. T. Macaluso, and J. Y. Chan, Phys. Rev. Lett. {\bf99}, 157203 (2007).
\bibitem{Myoung2008} B. R. Myoung, S. J. Kim, and C.S. Kim, J. Korean Phys. Soc.  {\bf53}, 750 (2008).
\bibitem{Zhao2012} S. Zhao, P. Dalmas de R\'{e}otier, A. Yaouanc, D. E. MacLaughlin, J. M. Mackie, O. O. Bernal, Y. Nambu, T. Higo, and S. Nakatsuji, Phys. Rev. B  {\bf86}, 064435 (2012).
\bibitem{Stock2010} C. Stock, S. Jonas, C. Broholm, S. Nakatsuji, Y. Nambu, K. Onuma, Y. Maeno, and J. H. Chung, Phys. Rev. Lett. {\bf105}, 037402 (2010).
\bibitem{Reotier2012} P. Dalmas de R\'{e}otier, A. Yaouanc, D. E. MacLaughlin, S. Zhao, T. Higo, S. Nakatsuji, Y. Nambu, C. Marin, G. Lapertot, A. Amato, and C. Baines, Phys. Rev. B {\bf85}, 140407(R) (2012). 
\bibitem{Nakatsuji2005} S. Nakatsuji, Y. Nambu, H. Tonomura, O. Sakai, S. Jonas, C. Broholm, H. Tsunetsugu, Y. Qiu, and Y. Maeno, Science  {\bf309}, 1697 (2005).

\bibitem{Mazin2007} I. I. Mazin, Phys. Rev. B  {\bf76}, 140406(R) (2007).
\bibitem{Tsunetsugu2006} H. Tsunetsugu and M. Arikawa, J. Phys. Soc. Jpn {\bf75}, 083701 (2006).
\bibitem{Lauchli2006} A. L\"{a}uchli, F. Mila, and K. Penc, Phys. Rev. Lett. {\bf97}, 087205 (2006).
\bibitem{Bhattacharjee2006} S. Bhattacharjee, V. B. Shenoy, and T. Senthil, Phys. Rev. B {\bf74}, 092406 (2006).
\bibitem{Valentine2020} M. E. Valentine, T. Higo, Y. Nambu, D. Chaudhuri, J. Wen, C. Broholm, S. Nakatsuji, and N. Drichko, Phys. Rev. Lett.  {\bf125}, 197201 (2020).
\bibitem{Tsunetsugu2007} H. Tsunetsugu and M. Arikawa, J. Phys.: Condens. Matter {\bf19},145248 (2007). 
\bibitem{Podolsky2009} D. Podolsky and Y.B. Kim, Phys. Rev. B  {\bf79}, 140402(R) (2009).
\bibitem{note1} Details of sample preparation and experimental setups are documented in Appendix \ref{Sec6}.
\bibitem{Takeya2008} H. Takeya, K. Ishida, K. Kitagawa, Y. Ihara, K. Onuma, Y. Maeno, Y. Nambu, S. Nakatsuji, D. E. MacLaughlin, A. Koda, and R. Kadono, Phys. Rev. B {\bf77}, 054429 (2008).
\bibitem{Nambu2009} Y. Nambu, R.T. Macaluso, T. Higo, K. Ishida, and S. Nakatsuji, Phys. Rev. B  {\bf79}, 214108 (2009).
\bibitem{Warren1941} B. E. Warren,  Phys. Rev. {\bf59}, 693 (1941).
\bibitem{Paddison2013} J. A. M. Paddison, J. R. Stewart, and A. L. Goodwin, J. Phys.: Condens. Matter {\bf25},454220 (2013). 

\bibitem{Tamura2008} R. Tamura and N. Kawashima, J. Phys. Soc. Jpn {\bf77}, 103002 (2008).
\bibitem{Tamura2011} R. Tamura and N. Kawashima, J. Phys. Soc. Jpn {\bf80}, 074008 (2011).
\bibitem{SpinW} https://spinw.org
\bibitem{Schrenk2014} K. J. Schrenk, N. A. M. Ara\'{u}jo, and H. J. Herrmann, Phys. Rev. E  {\bf87}, 032123 (2013).
\bibitem{Halperin1977} B. I. Halperin and W. M. Saslow, Phys. Rev. B  {\bf16}, 2154 (1977).
\bibitem{Ajiro1988} Y. Ajiro, H. Kikuchi, S. Sugiyama, T. Nakashima, S. Shamoto, N. Nakayama, M. Kiyama, N. Yamamoto, and Y. Oka, J. Phys. Soc. Jpn {\bf57}, 2268 (1988).
\bibitem{Wollny2011} A. Wollny, L. Fritz, and M. Vojta, Phys. Rev. Lett. {\bf107}, 137204 (2011).
\bibitem{Maryasin2013} V. S. Maryasin and M. E. Zhitomirsky, Phys. Rev. Lett. {\bf111}, 247201 (2013).
\bibitem{fullprof1993} J. Rodr\'{i}guez-Carvajal, Phys. B: Condens. Matt. {\bf192}, 55 (1993).
\bibitem{shelx} G. M. Sheldrick, Acta Cryst. {\bf {C71}}, 3(2015).
\bibitem{Stewart2008} J. R. Stewart, P. P. Deen, K. H. Andersen, H. Schober, J. F. Barthelemy, J. M. Hillier, A. P. Mjurani, T. Hayes, and B. Lindenau, J Appl. Cryst.  {\bf42}, 69 (2009).
\bibitem{BewleyLET} R. I. Bewley, J. W. Taylor, S. M. Bennington, Nucl. Instrum. Methods Phys. Res. Sect. A {\bf637}, 128 (2011). 
\bibitem{Yamaguchii2005} H. Yamaguchi, S. Kimura, M. Hagiwara, Y. Nambu, S. Nakatsuji, Y. Maeno,  A. Matsuo, and K. Kindo, J. Phys. Soc. Jpn {\bf79}, 054710 (2010).
\end{thebibliography}
\end{document}